\documentclass[english,showpacs,preprint,superscriptaddress,prl]{revtex4-1}
\usepackage{lmodern}
\usepackage{lmodern}
\usepackage[T1]{fontenc}
\usepackage{amsmath}
\usepackage{amssymb}
\usepackage{graphicx}
\usepackage{esint}

\makeatletter

\usepackage{amsthm}\usepackage{latexsym}\usepackage{bm}\usepackage{amsfonts}

\usepackage{babel}

\usepackage{hyperref}
\hypersetup{
	breaklinks = true,
    colorlinks = true,
    citecolor = {blue},
	urlcolor = {blue},
	linkcolor = {blue}
}

\usepackage{babel}

\global\long\def\EXP{\times10^}
\usepackage{cleveref}

 \global\long\def\rmd{\mathrm{d}}

 \global\long\def\bfx{\mathbf{x}}
 \global\long\def\bfX{\mathbf{X}}
 \global\long\def\bfxz{\mathbf{x}_{l-1}}
 \global\long\def\bfxo{\mathbf{x}_{l}}
 \global\long\def\bfxt{\mathbf{x}_{l+1}}
 \global\long\def\bfp{\mathbf{p}}
 \global\long\def\bfv{\mathbf{v}}

 \global\long\def\bfA{\mathbf{A}}
 \global\long\def\bfb{\mathbf{b}}
 
 \global\long\def\bfB{\mathbf{B}}

 \global\long\def\bfE{\mathbf{E}}

 \global\long\def\bfe{\mathbf{e}}

 \global\long\def\FIG#1{Fig.~\ref{#1}}
 \global\long\def\EQ#1{Eq.~(\ref{#1})}

\usepackage{babel}

\@ifundefined{showcaptionsetup}{}{%
 \PassOptionsToPackage{caption=false}{subfig}}
\usepackage{subfig}
\makeatother

\usepackage{babel}
\begin{document}

\title{Slow manifolds of classical Pauli particle enable structure-preserving
geometric algorithms for guiding center dynamics}

\author{Jianyuan Xiao}
\email{xiaojy@ustc.edu.cn}

\selectlanguage{english}%

\affiliation{Department of Engineering and Applied Physics, University of Science
and Technology of China, Hefei, 230026, China}

\author{Hong Qin}
\email{hongqin@princeton.edu}

\selectlanguage{english}%

\affiliation{Plasma Physics Laboratory, Princeton University, Princeton, NJ 08543,
U.S.A.}
\begin{abstract}
Since variational symplectic integrators for the guiding center was
proposed \citep{qin2008variational,qin2009variational}, structure-preserving
geometric algorithms have become an active research field in plasma
physics. We found that the slow manifolds of the classical Pauli particle
enable a family of structure-preserving geometric algorithms for guiding
center dynamics with long-term stability and accuracy. This discovery
overcomes the difficulty associated with the unstable parasitic modes
for variational symplectic integrators when applied to the degenerate
guiding center Lagrangian. It is a pleasant surprise that Pauli's
Hamiltonian for electrons, which predated the Dirac equation and marks
the beginning of particle physics, reappears in classical physics
as an effective algorithm for solving an important plasma physics
problem. This technique is applicable to other degenerate Lagrangians
reduced from regular Lagrangians. 
\end{abstract}



\maketitle
Guiding center dynamics lies at the heart of the gyrokinetic theory.
The success of widely adopted gyrokinetic simulations depends on the
effectiveness of the algorithms for numerically integrating the guiding
center dynamics. Standard integration algorithms for differential
equations, such as the Runge-Kutta methods, do not preserve the geometric
structure of the guiding center dynamics, and the truncation error
from each time-step accumulates coherently. As a consequence, long-term
simulation results by these standard algorithms are not trustworthy.
A solution to this problem proposed in 2008 \citep{qin2008variational,qin2009variational}
is to design a symplectic variational algorithm for the guiding center.
This idea has grown into an active research field of structure-preserving
geometric algorithms for plasma physics \citep{squire4748,squire2012geometric,xiao2013variational,xiao2015explicit,xiao2015variational,he2015hamiltonian,qin2016canonical,he2016hamiltonian,kraus2017gempic,xiao2017local,xiao2018structure,xiao2019field,Xiao2020,Glasser2020}.
These new algorithms have been successfully applied to study important
physics problems that are otherwise difficult to simulate using conventional
algorithms. Examples include whole-device 6D kinetic simulations of
tokamak physics \citep{Xiao2020}, numerical confirmation \citep{qin2016canonical}
of Mouhot and Villani's theory on nonlinear Landau damping \citep{Mouhot2011},
first-principles based real-time lattice simulation of quantum plasmas
\citep{shi2018}, and the strongest numerical evidence in support
of Paker's conjecture of singular current formation \citep{Zhou2017APJ}. 

Because the symplectic structure of the guiding center is non-canonical,
standard canonical symplectic integrators \citep{ruth83,feng85,feng86,SanzSerna1988,feng10,sanz-serna94,hairer02}
are not applicable. The non-canonical symplectic integrators for the
guiding center first proposed \citep{qin2008variational,qin2009variational}
are based on the discrete variational principle \citep{marsden2001discrete}.
It was soon realized \citep{Shang2011,ellison2015,lelandthesis} that
due to the degenerate nature of the guiding center Lagrangian, the
algorithms are two-step methods, which introduce extra parasitic modes
to the discrete systems. These parasitic modes could lead to numerical
instability in certain parameter regimes. A few remedies have been
proposed using the methods of canonicalization \citep{zhang2014canonicalization},
regularization \citep{burby2017}, projection \citep{kraus2017},
or degeneracy \citep{ellison2018}. However, these methods are subject
to various restrictions. A practical symplectic integrator for guiding
centers in general magnetic fields remains elusive. 

In the paper, we present a family of structure-preserving geometric
algorithms with long-term stability and accuracy for the guiding center
dynamics based on the slow manifold dynamics of the Classical Pauli
Particle (CPP). Historically, Pauli's Hamiltonian for electrons predated
the Dirac Hamiltonian, and marks the beginning of particle physics.
It is a serendipity that the physics of the classical Pauli particle
solves a challenge in computational plasma physics. The algorithms
are valid for arbitrary magnetic fields and can be directly implemented
using the standard laboratory phase space coordinates. Numerical experiments
have confirmed the long-term stability and accuracy of the algorithms.

We start our algorithm design by considering the Lagrangian of the
classical Pauli particle with a scalar magnetic moment $\mu,$ 
\begin{equation}
L_{\mathrm{cpp}}\left(\bfx,\dot{\bfx}\right)=\frac{1}{2}\left|\dot{\bfx}\right|^{2}+\left(\dot{\bfx}\cdot\bfA\left(\bfx\right)-\phi\left(\bfx\right)-\mu B\left(\bfx\right)\right)\,.\label{eq:Lcpp}
\end{equation}
The mass and charge of the particle are set to $1.$ The corresponding
Hamiltonian in the canonical coordinates is 
\begin{equation}
H_{\mathrm{cpp}}\left(\bfp,\bfx\right)=\frac{1}{2}\left(\bfp-\bfA\left(\bfx\right)\right)^{2}+\mu B\left(\bfx\right)+\phi\left(\bfx\right)\,,\label{eq:Hcpp}
\end{equation}
where $\bfp=\bfA+\dot{\bfx}~$ is the canonical momentum. The novelty
of this Lagrangian and Hamiltonian is the inclusion of the magnetic
moment $\mu$. This particle is called classical Pauli particle because
its Hamiltonian is the classical version of Pauli's Hamiltonian for
electrons, 

\begin{equation}
\mathrm{H}_{\mathrm{Pauli}}=\frac{1}{2}\left(\bfp-\bfA\left(\bfx\right)\right)^{2}-\frac{\hbar}{2}\mathbf{\sigma}\cdot\mathbf{B}\left(\bfx\right)+\phi\left(\bfx\right)\,,
\end{equation}
where $\mathbf{\sigma}=(\mathbf{\sigma}_{x},\mathbf{\sigma}_{y},\mathbf{\sigma}_{z})$
is the vector of $2\times2$ Pauli matrices and the charge of electron
is $-1$. Pauli had introduced the intrinsic spin operators to explain
the intrinsic magnetic moment of electrons observed in the Stern--Gerlach
experiment, before Dirac wrote down his equation for electrons. In
most regimes of classical physics, the intrinsic magnetic moment of
charged particles is negligible. However, our study reveals that introducing
a formal magnetic moment term $\mu B$ in the Hamiltonian for classical
particles surprisingly leads to a family of structure-preserving geometric
algorithms for the guiding center dynamics. Given the fundamental
importance of Paul's Hamiltonian in particle physics, this discovery
should not be totally surprising except that the utility of Pauli's
Hamiltonian in classical physics is manifested as an effective algorithm
for solving an important plasma physics problem.

Unlike the guiding center Lagrangian, the Lagrangian for the CPP $L_{\mathrm{cpp}}$
is regular, and many known structure-preserving geometric algorithms,
including those custom-designed for classical charged particles \citep{qin2013boris,he2015hamiltonian,zhang2015volume,he2015volume,he2016high,he2016higher,zhang2016explicit,tu2016,tao2016,he2017explicit,zhou2017explicit,xiao2019a,shi2019},
can be directly applied. To simulate the guiding center dynamics,
we adopt a structure-preserving geometric algorithm and select an
initial condition such that $\dot{\bfx}\times\mathbf{b}=0$. Before
discussing the choices of structure-preserving geometric algorithms,
let's explain why this algorithm solves for the guiding center dynamics.
The guiding center Lagrangian is 
\begin{equation}
L_{\mathrm{gc}}\left(\bfX,\dot{\bfX},u,\dot{u}\right)=\left(\bfA\left(\bfX\right)+u\bfb\left(\bfX\right)\right)\cdot\dot{\bfX}-\left(\frac{1}{2}u^{2}+\mu B\left(\bfX\right)+\phi\left(\bfX\right)\right)\,,\label{EqnGC}
\end{equation}
where $\bfX$ is the guiding center and $u$ is the parallel velocity.
It is derived by Littlejohn \citep{littlejohn1983variational} using
a Lie perturbation method under the strong field ordering from the
standard Lagrangian for the classical particle,
\begin{equation}
L_{\mathrm{cp}}\left(\bfx,\dot{\bfx}\right)=\frac{1}{2}\left|\dot{\bfx}\right|^{2}+\left(\dot{\bfx}\cdot\bfA\left(\bfx\right)-\phi\left(\bfx\right)\right)\,.
\end{equation}
The variational symplectic integrators first proposed \citep{qin2008variational,qin2009variational}
were based on the discrete version of $L_{\mathrm{gc}}$. If we carry
out the same perturbative analysis to the Lagrangian of the CPP $L_{\mathrm{cpp}}$,
we will obtain the following guiding center Lagrangian for the CPP,
\begin{equation}
L_{\mathrm{cpp-gc}}\left(\bfX,\dot{\bfX},u,\dot{u}\right)=\left(\bfA\left(\bfX\right)+u\bfb\left(\bfX\right)\right)\cdot\dot{\bfX}-\left(\frac{1}{2}u^{2}+\mu'B\left(\bfX\right)+\mu B\left(\bfX\right)+\phi\left(\bfX\right)\right)\,.\label{EqnMGC}
\end{equation}
Here, $\mu'$ is the magnetic moment of the CPP associated with its
perpendicular kinetic energy, 
\begin{equation}
\mu'\approx\frac{|\dot{\bfx}\times\bfb|^{2}}{2B}\,.
\end{equation}
Comparing Eqs.\,(\ref{EqnMGC}) and (\ref{EqnGC}), we observe that
the only difference is the $\mu'B$ term in $L_{\mathrm{cpp-gc}}$.
If we set $\dot{\bfx}\times\bfb=0$ at $t=0$, then we will have $\mu'\approx0$
for a very long time \citep{Qin2006} because it is an adiabatic invariant.
And the location of the CPP should be nearly identical to its guiding
center since the gyro-radius is close to $0$, i.e., $\bfx\approx\bfX$.
Furthermore, when $\mu'\approx0$, we have $L_{\mathrm{\mathrm{gc}}}\approx L_{\mathrm{cpp-\mathrm{gc}}}$.
Therefore, the CPP will be very close to the guiding center of the
classical particle governed by the Lagrangian $L_{\mathrm{gc}}$.
The solutions of the CPP with $\dot{\bfx}\times\bfb\approx0$ for
a very long time can be viewed as slow manifolds \citep{Lorenz_1986,MacKay2004}
of the CPP dynamics. From this perspective, the guiding center dynamics
can be identified with the slow manifolds of the CPP. We note that
this viewpoint is similar to Burby's recent theory of guiding centers
as slow manifolds of loop dynamics \citep{Burby2020}. 

We now give three structure-preserving geometric integrators with
excellent long-term stability and accuracy for the slow manifolds
of the CPP, the gauge-independent symplectic algoritm, the midpoint
variational symplectic algoritm, and the volume preserving algorithm.

For the gauge-independent symplectic algorithm, a gauge-independent
discretization of $L_{\mathrm{cpp}}$ should be used. In the present
work, we adopt the following 2nd-order discrete action using a technique
similar to that in Ref.\,\citep{squire2012geometric},
\begin{eqnarray}
S_{\rmd} & = & \sum_{l}L_{\rmd}\left(\bfxz,\bfxo\right)\Delta t\,,\\
L_{d}\left(\bfxz,\bfxo\right) & = & \frac{1}{2}\left(\frac{\bfxo-\bfxz}{\Delta t}\right)^{2}+\frac{\bfxo-\bfxz}{\Delta t}\cdot\int_{0}^{1}\rmd\tau\bfA\left(\bfxz+\tau\left(\bfxo-\bfxz\right)\right)-\nonumber \\
 &  & \phi\left(\bfxo\right)-\mu B\left(\bfxo\right)\,.
\end{eqnarray}
The corresponding discrete Euler-Lagrangian (EL) equation is 
\begin{eqnarray}
\frac{\partial S_{\rmd}}{\partial\bfxo} & = & 0\,,
\end{eqnarray}
or more specifically, 
\begin{eqnarray}
\frac{\bfxt-2\bfxo+\bfxz}{\Delta t^{2}} & = & \bfE^{\dagger}\left(\bfxo\right)+\frac{\bfxo-\bfxz}{\Delta t}\times\int_{0}^{1}\rmd\tau\tau\bfB\left(\bfxz+\tau\left(\bfxo-\bfxz\right)\right)+\nonumber \\
 &  & \frac{\bfxt-\bfxo}{\Delta t}\times\int_{0}^{1}\rmd\tau\tau\bfB\left(\bfxt+\tau\left(\bfxo-\bfxt\right)\right)\,,\label{EqnDISCPUS}
\end{eqnarray}
where 
\begin{eqnarray}
\bfE^{\dagger} & = & -\nabla\left(\phi+\mu B\right)
\end{eqnarray}
is the modified electric field. This scheme is implicit since the
right-hand side of \EQ{EqnDISCPUS} also contains $\bfxt$, and
we can use Newton's method to solve for $\bfxt$. Compared with previous
works on geometric guiding center integrators \citep{qin2009variational,qin2008variational,li2011variational,zhang2014canonicalization,ellison2015,ellison2018},
the above algorithm enjoys several advantages. I) It is electromagnetic
gauge-free. This can be seen from the discrete EL equation (\ref{EqnDISCPUS}),
which depends only on electromagnetic fields $\mathbf{E}$ and $\mathbf{B}$.
The gauge-free property is more desirable in particle-in-cell methods
since it relates directly to the local charge conservation law \citep{squire2012geometric,xiao2015variational,kraus2017gempic,xiao2018structure,Glasser2020}.
II) The present scheme can be applied to general magnetic fields and
it needs neither canonicalization \citep{zhang2014canonicalization}
nor specific gauge transformation. III) Since it is based on a regular
Lagrangian, instead of a degenerate one, it is not a multi-step method
\citep{Shang2011,ellison2015,ellison2018}, and not subject to the
unstable parasitic modes. 

The second algorithm is the midpoint variational symplectic integrator
based on the following midpoint discrete action integral \citep{li2011variational,ellison2015,ellison2018},
\begin{eqnarray}
S_{\mathrm{dv}} & = & \sum_{l}L_{\mathrm{dv}}\left(\bfxz,\bfxo\right)\Delta t\,,\\
L_{\mathrm{dv}}\left(\bfxz,\bfxo\right) & = & L_{\mathrm{cpp}}\left(\frac{\bfxz+\bfxo}{2},\frac{\bfxo-\bfxz}{\Delta t}\right)\,.
\end{eqnarray}
The corresponding iteration rule is again by the discrete EL equation
\begin{eqnarray}
\frac{\partial S_{\mathrm{dv}}}{\partial\bfxo}=0\,.\label{EqnVSI}
\end{eqnarray}
Given $\bfxz$ and $\bfxo$, $\bfxt$ can be solved for from \EQ{EqnVSI}.
Its main advantage compared with the first algorithm is that it does
not require calculating integrals. In practice, it runs much faster
than the first algorithm when these integrals are expensive to evaluate.
The drawback is that it is not electromagnetic gauge-free. According
to previous investigations \citep{suris1990hamiltonian,marsden2001discrete},
the variational integrator applied to this non-degenerate Lagrangian
is equivalent to a canonical symplectic partitioned Runge-Kutta method
applied to the Hamiltonian specified by Eq.\,(\ref{eq:Hcpp}). Thus,
we may also refer to this variational symplectic integrator as a canonical
symplectic integrator.

The third algorithm is the volume-preserving method based on the original
Boris algorithm \citep{boris70}. If we treat the additional $\mu B$
term in the CPP Lagrangian as an extra electric potential, then the
Boris algorithm can be directly applied, 
\begin{eqnarray}
\frac{\bfxo-\bfxz}{\Delta t} & = & \bfv_{l-1/2}\,,\\
\frac{\bfv_{l+1/2}-\bfv_{l-1/2}}{\Delta t} & = & \bfE^{\dagger}\left(\bfxo\right)+\frac{\bfv_{l+1/2}-\bfv_{l-1/2}}{2}\times\bfB\left(\bfxo\right)\,.
\end{eqnarray}
This is a scheme using $\bfxz$ and $\bfv_{l-1/2}$ to obtain $\bfxo$
and $\bfv_{l+1/2}$. According to previous investigations \citep{qin2013boris},
the above scheme is volume preserving and it also possesses a good
long-term conservation property as the symplectic integrators do in
many cases. Moreover, the Boris algorithm is explicitly solvable and
requires neither calculating integrals nor the knowledge of potentials,
which significantly reduces the computational cost. It is preferable
in test particle simulations. 

It is worth mentioning that recently He et al. developed a family
of explicit high-order noncanonical Hamiltonian splitting methods
\citep{he2015hamiltonian,he2017explicit,zhou2017explicit,xiao2019a}
and high-order volume preserving algorithms \citep{he2015volume,he2016high,he2016higher}
for the charged particle dynamics , and of course they can be applied
to solve the CPP dynamics. The Poisson brackets for the CPP and the
classical particle are the same, and both Hamiltonians are separable
and all subsystems admit analytical solutions or volume preserving
maps. However, since the algorithms are usually explicit, they may
not be able to bound small deviations from the slow manifolds when
the time-step is comparable or larger than the gyro-period. In the
present study, we have not investigated the applicability of these
algorithms for slow manifold dynamics.

The main advantage of numerical integration of the guiding centers
dynamics rather than the charged particle dynamics is that the time-step
for guiding center integrators can be much larger than the gyro-period.
We now use numerical experiments to demonstrate that the three structure-preserving
geometric algorithms listed above can correctly calculate the guiding
center dynamics as slow manifolds of the CPP using large time-steps
with long-term stability and accuracy. Six algorithms are tested.
For the slow manifold dynamics of the CPP, the three structure-preserving
geometric algorithms are the 2nd-order gauge-invariant implicit symplectic
method (GISIP2), the midpoint variational symplectic integrator (VSIP2),
and the Boris algorithm (BAP2). For comparison, the guiding center
dynamics is also simulated by the implicit midpoint variational symplectic
integrator applied to the guiding center Lagrangian $L_{\mathrm{gc}}$
(VSI2) , the 4th-order Runge-Kutta method applied to the guiding center
equation (RK4), and the original Boris algorithm applied to Newton's
equation of the classical particle (BA2). 

\begin{figure}[htp]
\begin{centering}
\subfloat[BA2]{\includegraphics[width=0.33\linewidth]{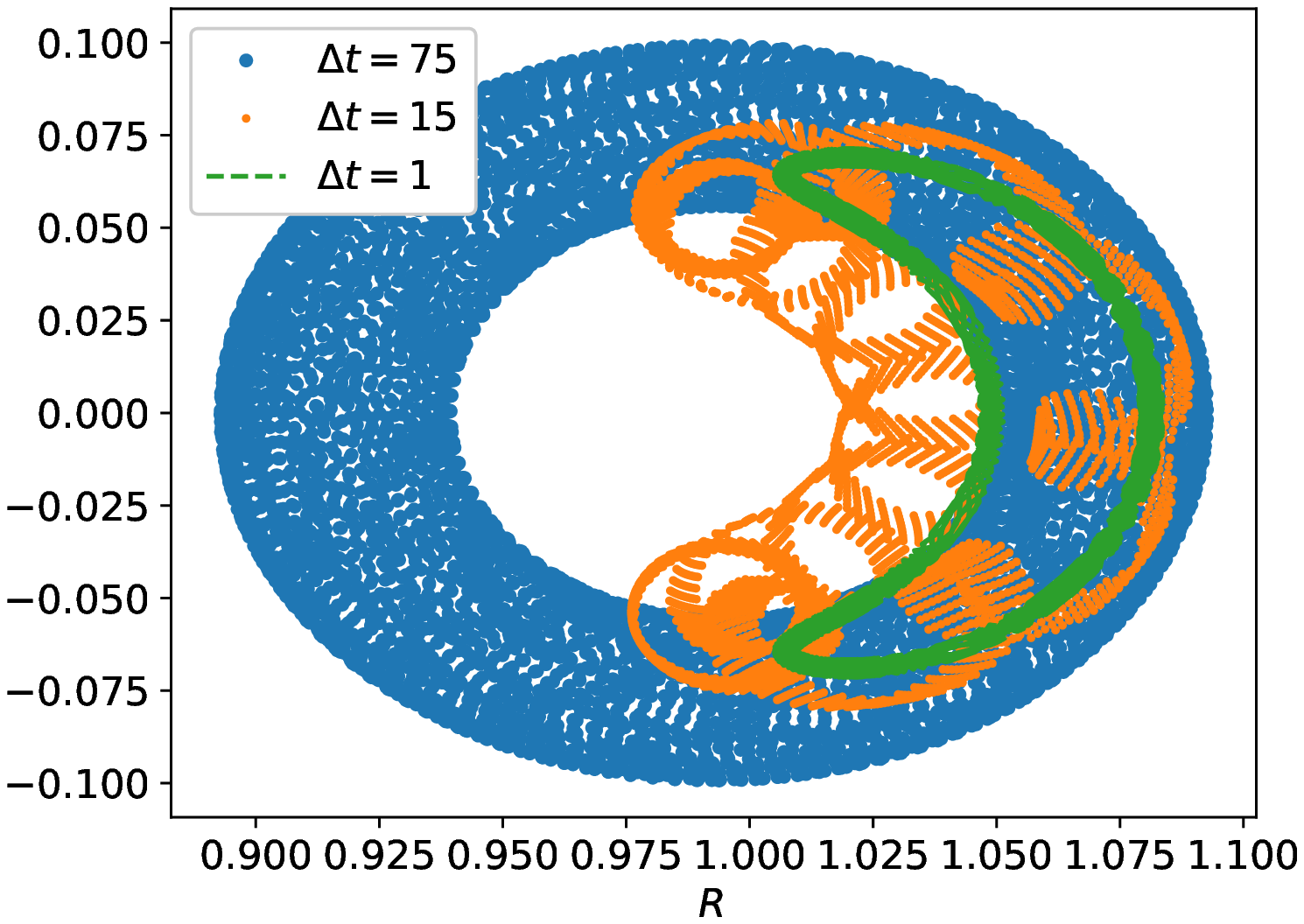}

}\subfloat[RK4]{\includegraphics[width=0.33\linewidth]{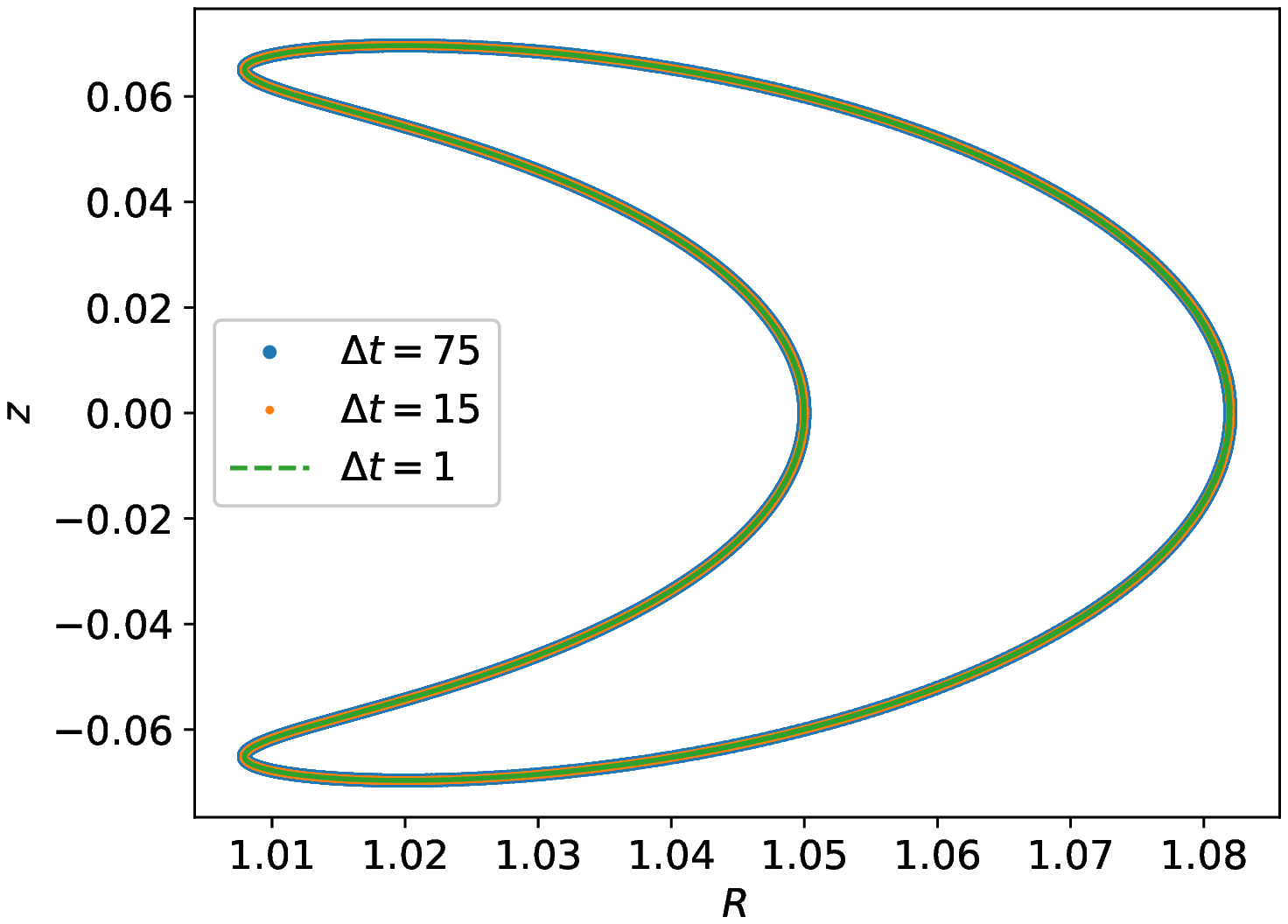}

}\subfloat[VSI2]{\includegraphics[width=0.33\linewidth]{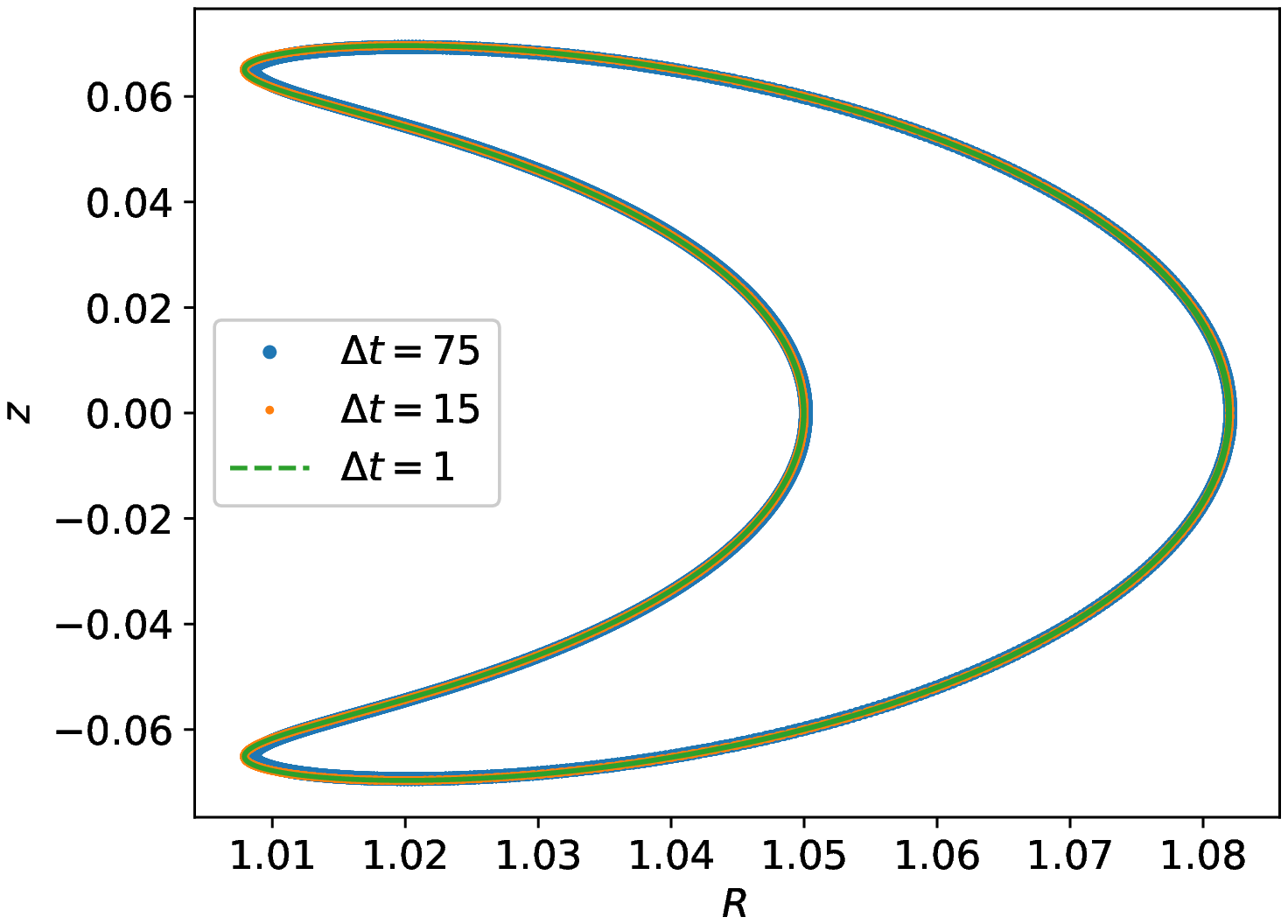}

}
\par\end{centering}
\begin{centering}
\subfloat[BAP2]{\includegraphics[width=0.33\linewidth]{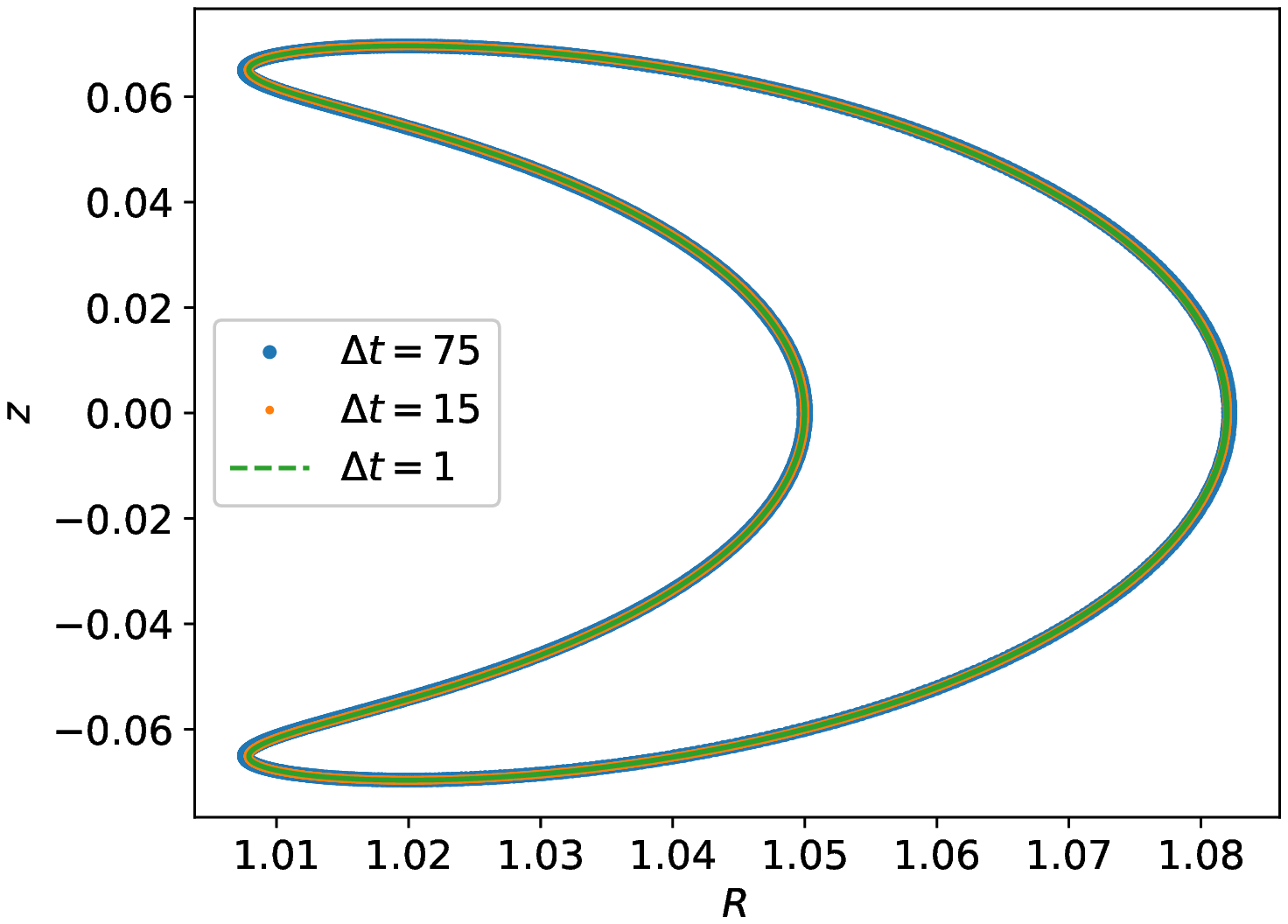}

}\subfloat[VSIP2]{\includegraphics[width=0.33\linewidth]{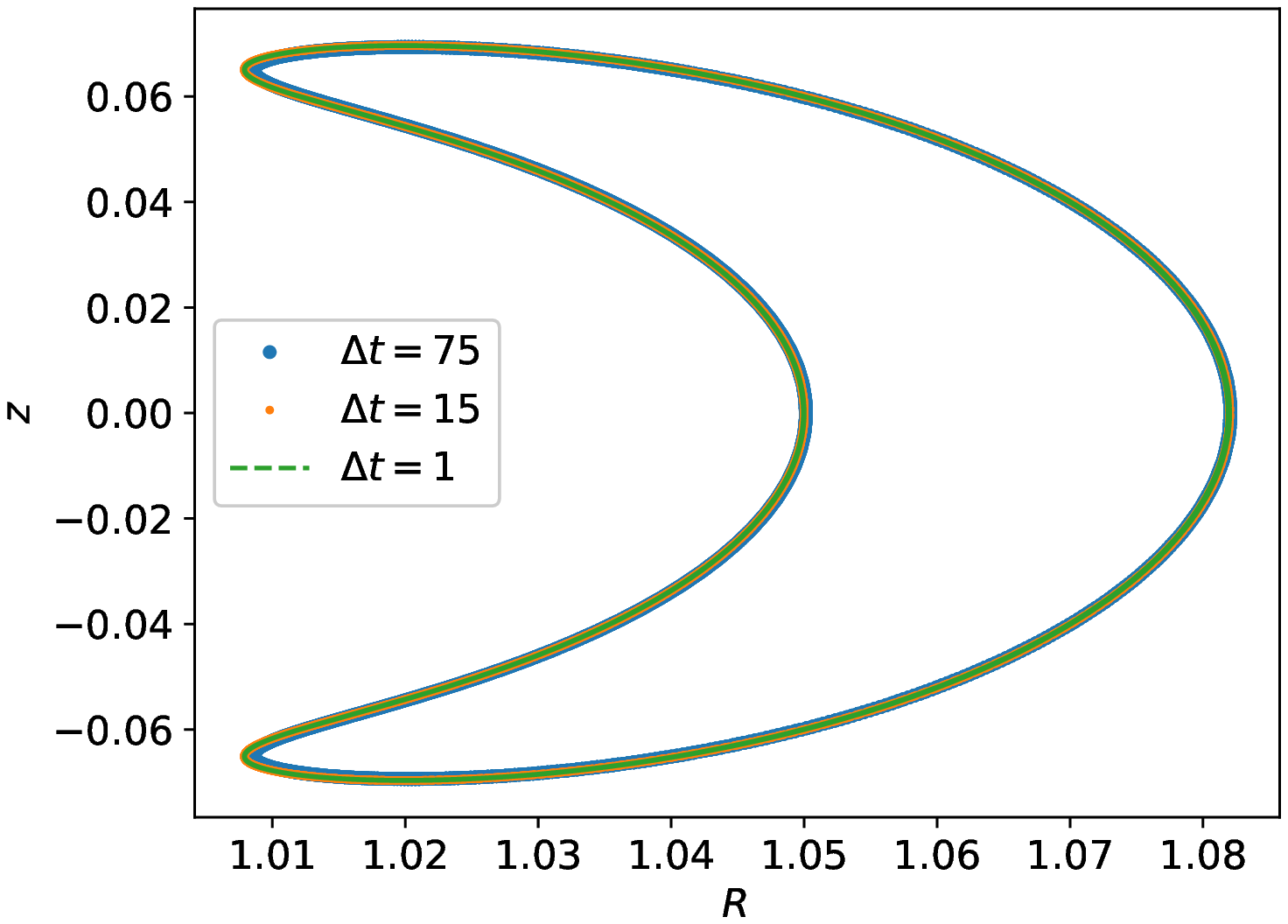}

}\subfloat[GISIP2]{\includegraphics[width=0.33\linewidth]{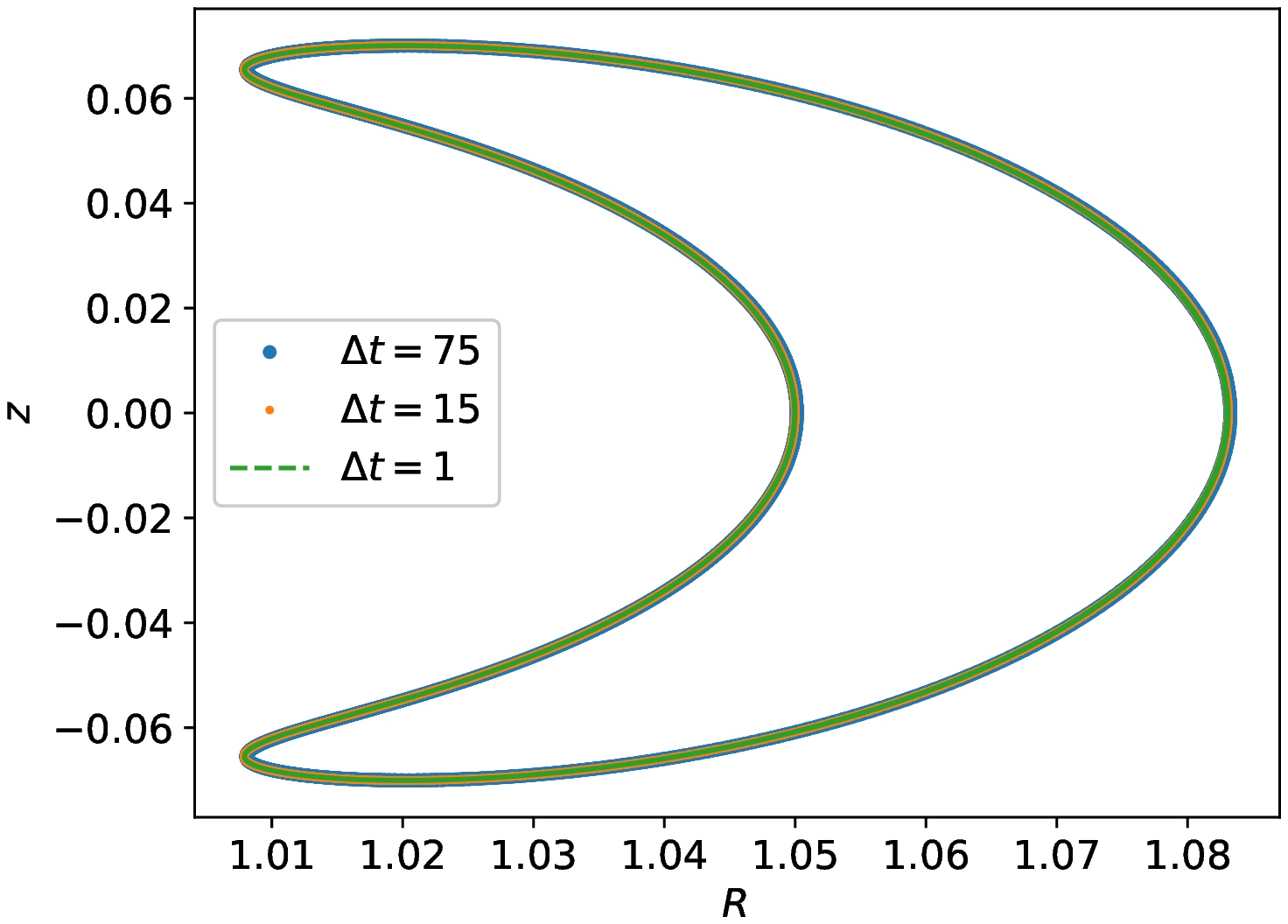}

}
\par\end{centering}
\caption{Banana orbits calculated by different algorithms with different time-steps.
All three structure-preserving geometric algorithms for the CPP dynamics
(BAP2, GISIP2 and VSIP2) can accurately calculate the banana orbit
as a slow manifold of the CPP with large time-steps. }
\label{FigBAE} 
\end{figure}

The numerical experiments are in the simplified tokamak field as described
in Ref.\,\citep{qin2009variational}. The potentials are 
\begin{eqnarray}
\bfA\left(x,y,z\right) & = & \frac{1}{2}B_{0}\left(\frac{r^{2}}{2R}\bfe_{\xi}-\log(R)\bfe_{z}+\frac{z}{2R}\bfe_{R}\right)\,,\\
\phi\left(x,y,z\right) & = & 0\,,
\end{eqnarray}
where 
\begin{eqnarray}
R & = & r\sqrt{x^{2}+y^{2}}\,,r=\sqrt{\left(R-1\right)^{2}+z^{2}}\,,\\
\bfe_{\xi} & = & \bfe_{R}[-\frac{y}{R},\frac{x}{R},0]\,,\bfe_{R}=[\frac{x}{R},\frac{y}{R},0]\,,
\end{eqnarray}
and $B_{0}=1$ is the strength of magnetic field at $R=1,z=0$. Initially
the particle's location and velocity are $\bfx_{0}=[1.05,0,0]$ and
$\bfv_{0}=[2.1\EXP{-3},4.3\EXP{-4},0],$ corresponding to initial
parallel velocity $u_{0}\approx4.3\EXP{-4}$ and magnetic moment $\mu\approx2.31\EXP{-6}$.
The gyro-period of the particle is approximately $2\pi$. First, we
test the six algorithms with different time-steps $\Delta t\in\left\{ 1,15,75\right\} $.
The total simulation time is $2\EXP{6}$. Simulated orbits on the
$R-z$ plane are plotted in the Fig. \ref{FigBAE}.

For this set of parameters, the particle is trapped, and its projection
on the $R-z$ plane is a banana orbit. When the time-step is larger
than the gyro-period, the original Boris algorithm applied to the
Newton's equation of the classical particle (BA2) gives incorrect
orbits. This indicates BA2 can not capture the slow drift motion in
the tokamak geometry using large time-steps. All other five algorithms
are stable at this time-scale and calculate the slow drift orbits
correctly. 

\begin{figure}[htp]
\begin{centering}
\subfloat[BAP]{\includegraphics[width=0.4\linewidth]{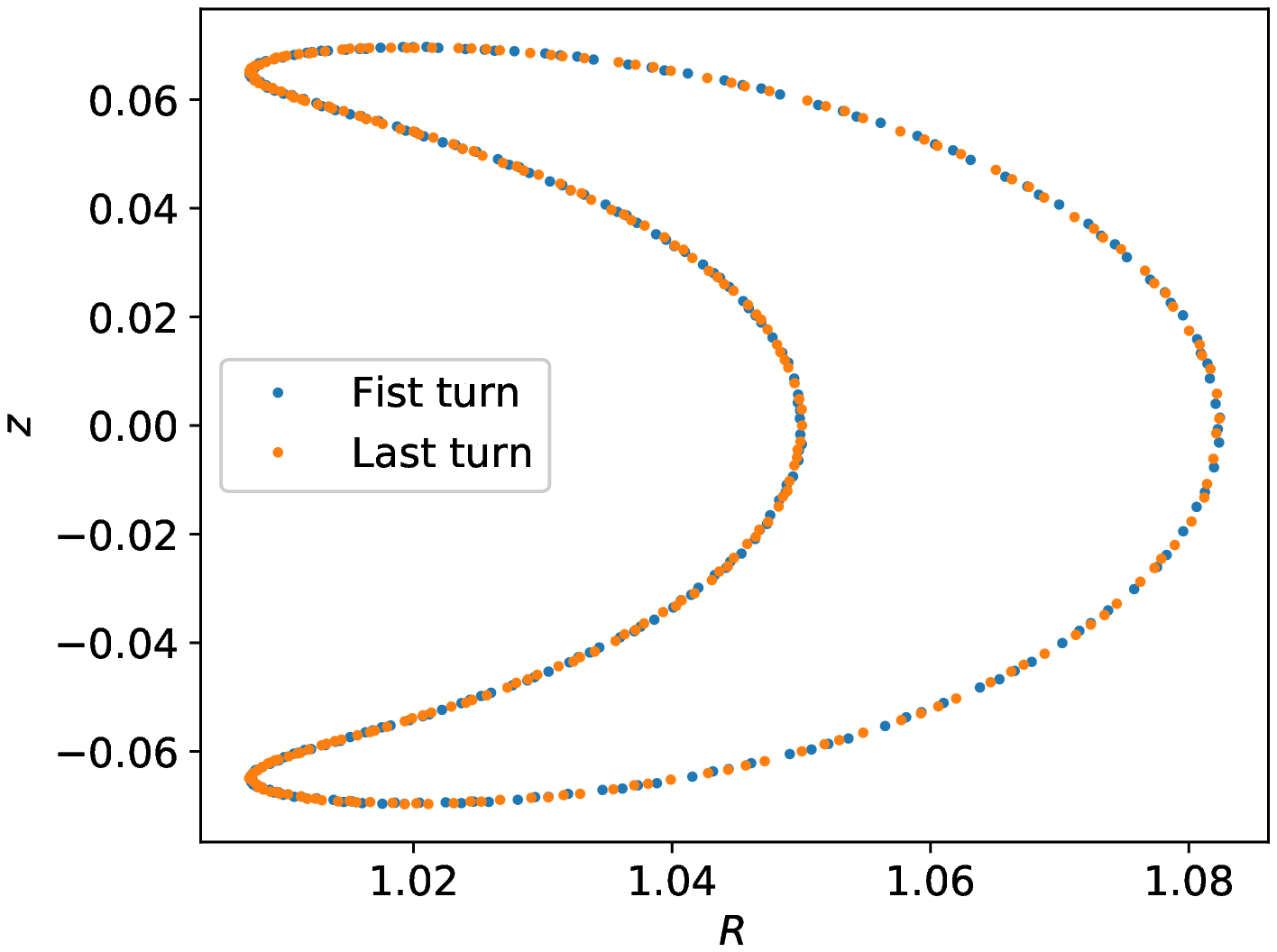}

}\subfloat[GISIP2]{\includegraphics[width=0.4\linewidth]{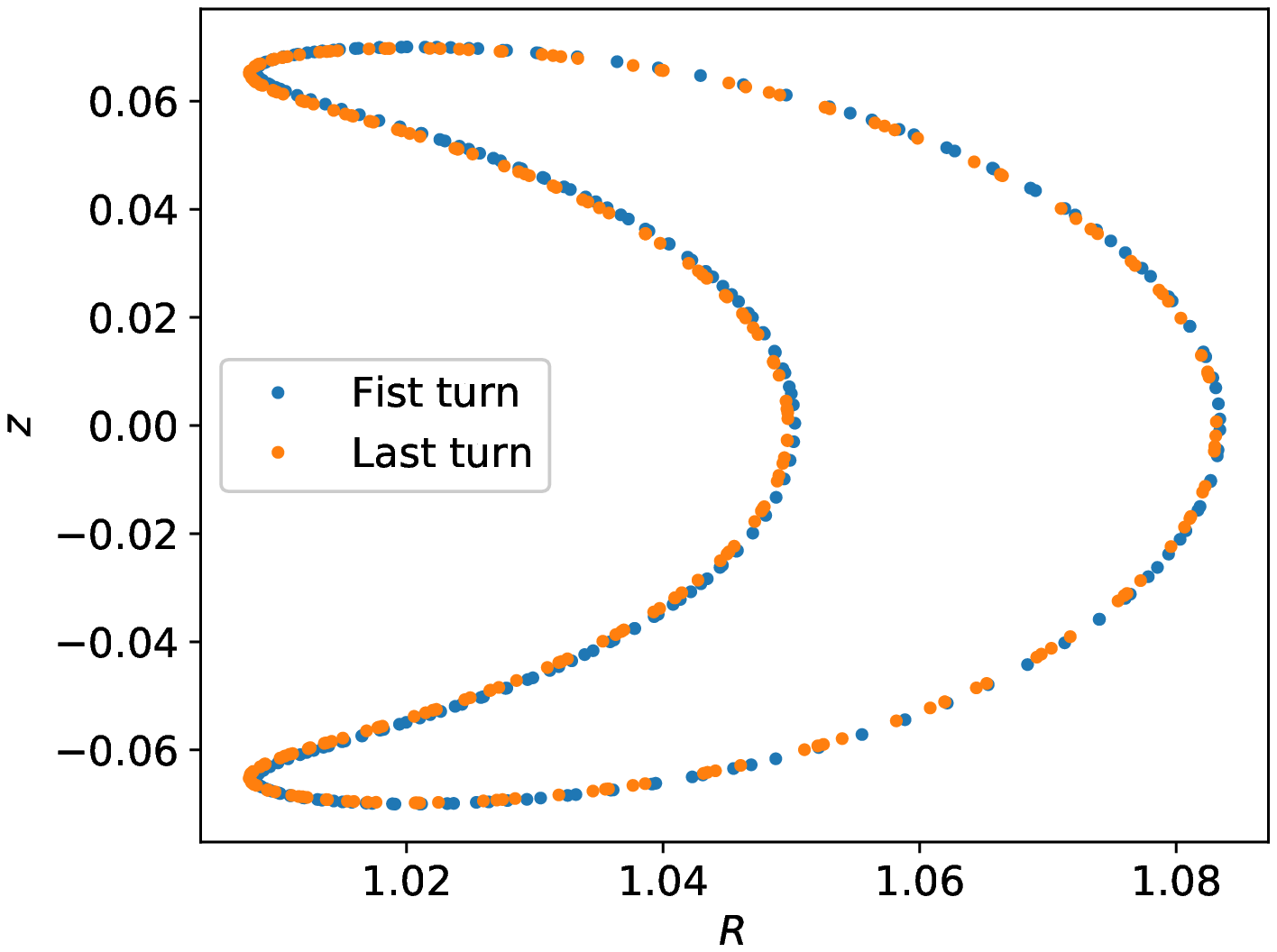}

}
\par\end{centering}
\begin{centering}
\subfloat[VSIP2]{\includegraphics[width=0.4\linewidth]{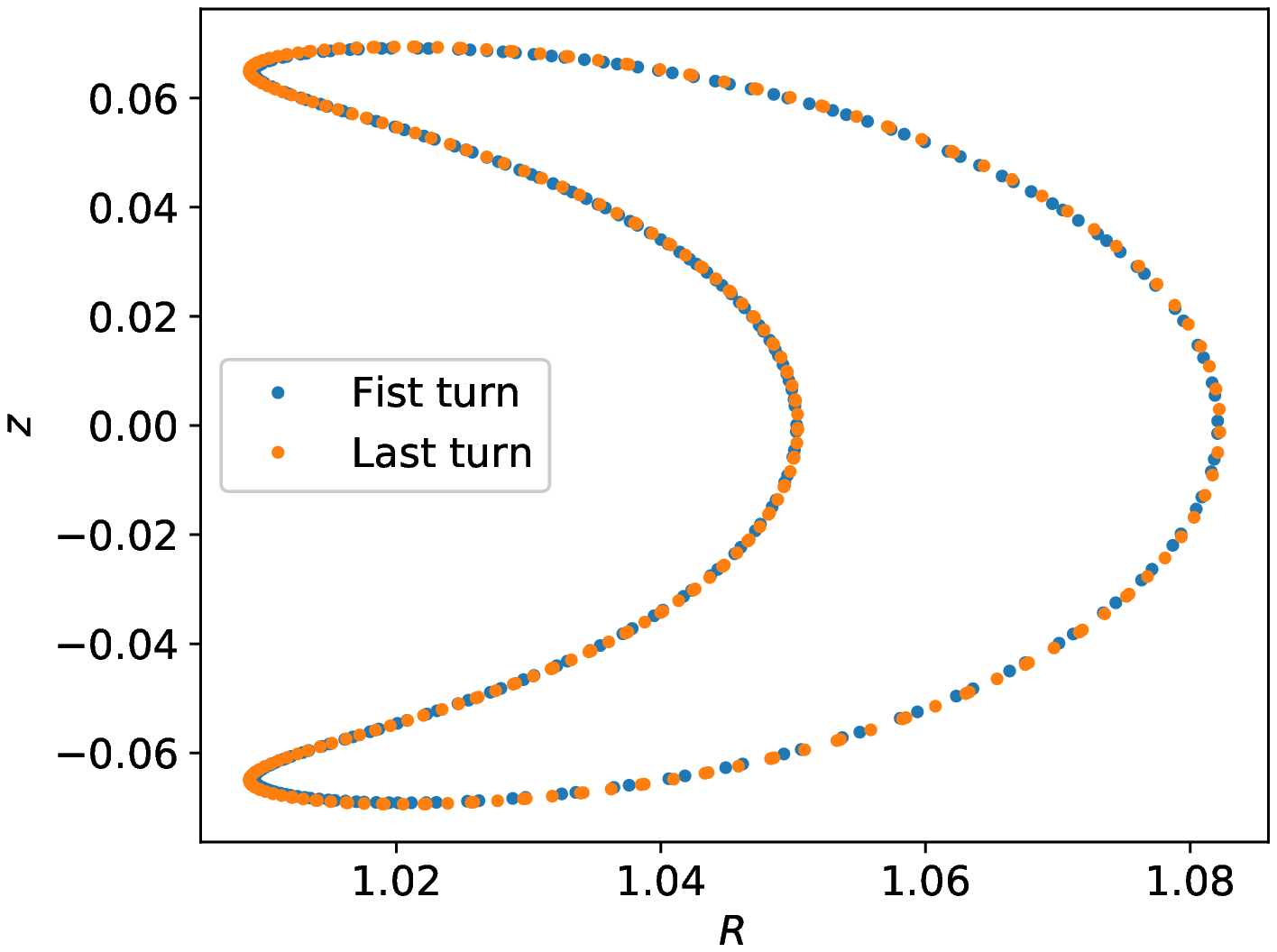}

}\subfloat[RK4]{\includegraphics[width=0.4\linewidth]{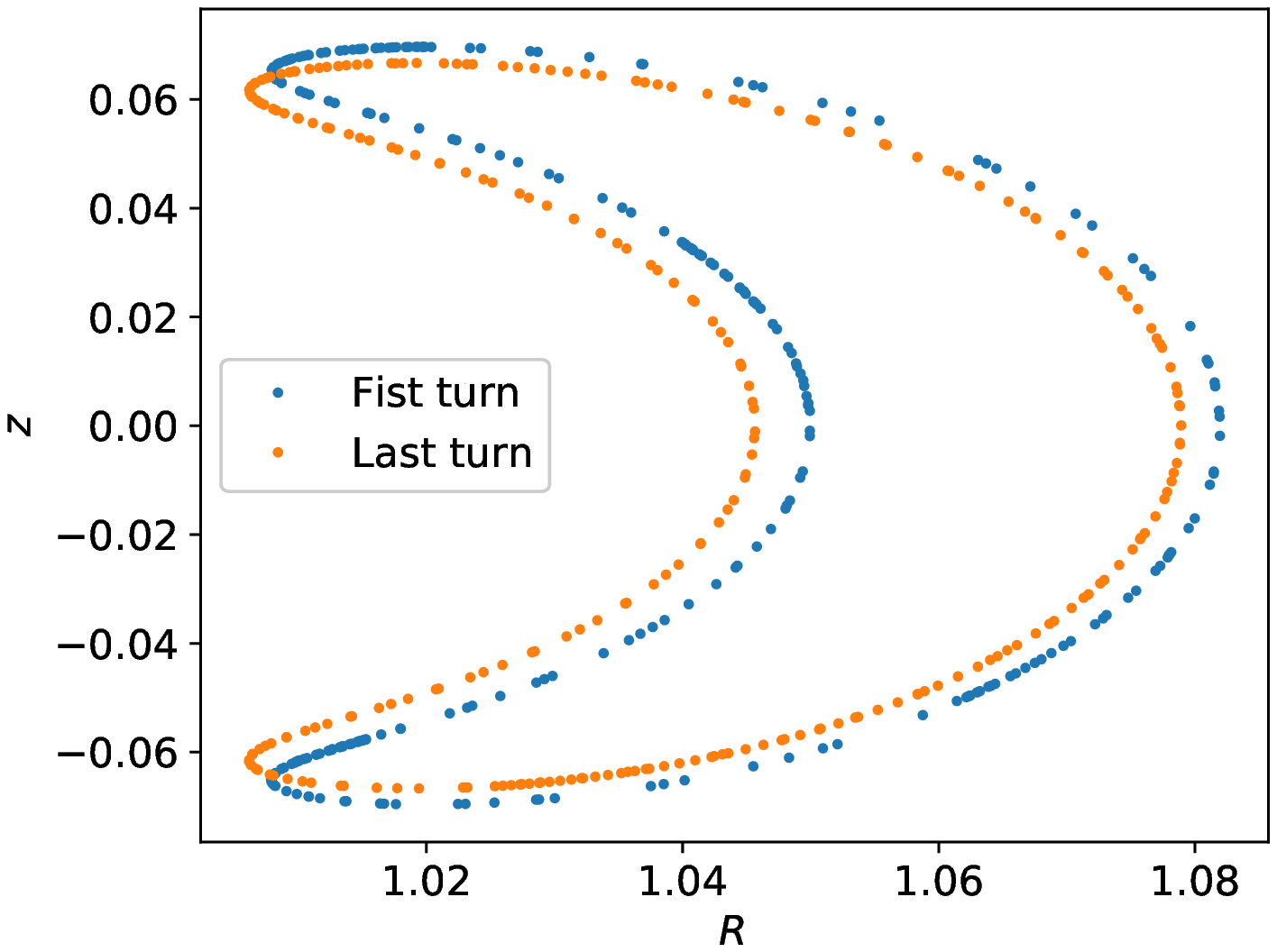}

}
\par\end{centering}
\caption{Comparison of long-term accuracy of the three structure-preserving
geometric algorithms (BAP2, GISIP2 and VSIP2) for the slow manifold
dynamics of the CPP and the RK4 method. The time-step is $\Delta t=105$.
The BAP2, GISIP2, and VSIP2 algorithms calculate the banana orbit
accurately for $1\EXP{8}$ time-steps. The energy error of the RK4
method accumulates coherently over time, and the long-terms simulation
result is not trustworthy. }
\label{Fig1051Y} 
\end{figure}

To demonstrate the long-term conservation property of structure-preserving
geometric algorithms for the slow manifold dynamics of the CPP, we
test the algorithms using a large time-step, i.e., $\Delta t=105$,
and run the simulations for $1\EXP{8}$ time-steps. The first and
last turns of the banana orbit in the poloidal plane calculated by
different algorithms are shown in \FIG{Fig1051Y}. It is clear that
all three structure-preserving geometric algorithms for the CPP dynamics
(BAP2, GISIP2 and VSIP2) can calculate the banana orbit accurately
as a slow manifold for $1\EXP{8}$ time-steps, while the non-geometric
RK4 algorithm can not. For the RK4 algorithm, the truncation error
from each time-step accumulates coherently, and the energy of the
discrete system monotonically decreases as a function of time. As
a result of this numerical dissipation, the banana orbit shrinks towards
the center of the device. This numerical error may mimic real physical
effects such as the neoclassical Ware pinch \citep{Ware_1970}. Without
long-term accuracy, the long-term simulation results of the RK4 method
are not reliable. 

\begin{figure}[htp]
\begin{centering}
\subfloat[$\Delta t=6$]{\includegraphics[width=0.4\linewidth]{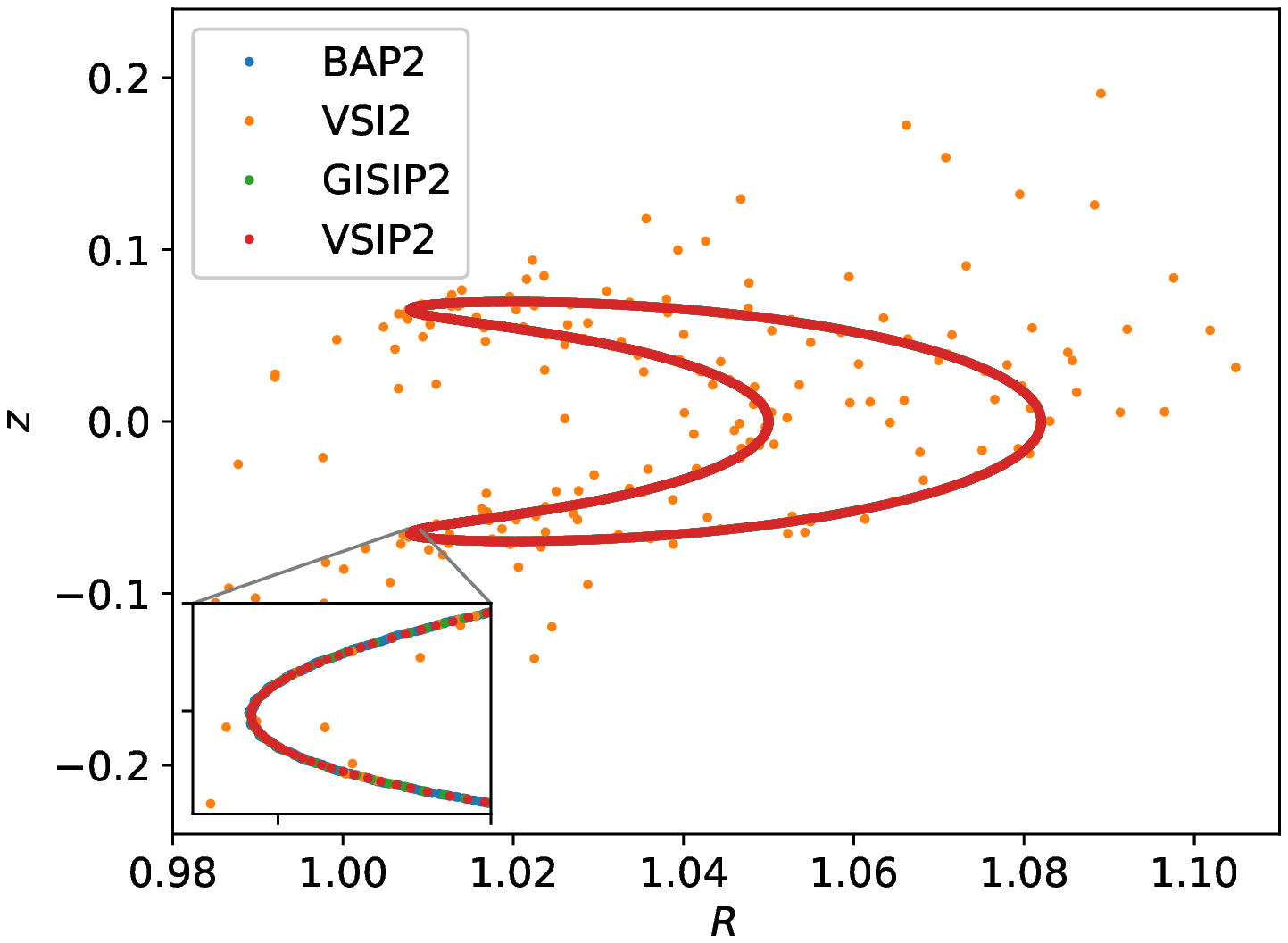}

}\subfloat[$\Delta t=15$]{\includegraphics[width=0.4\linewidth]{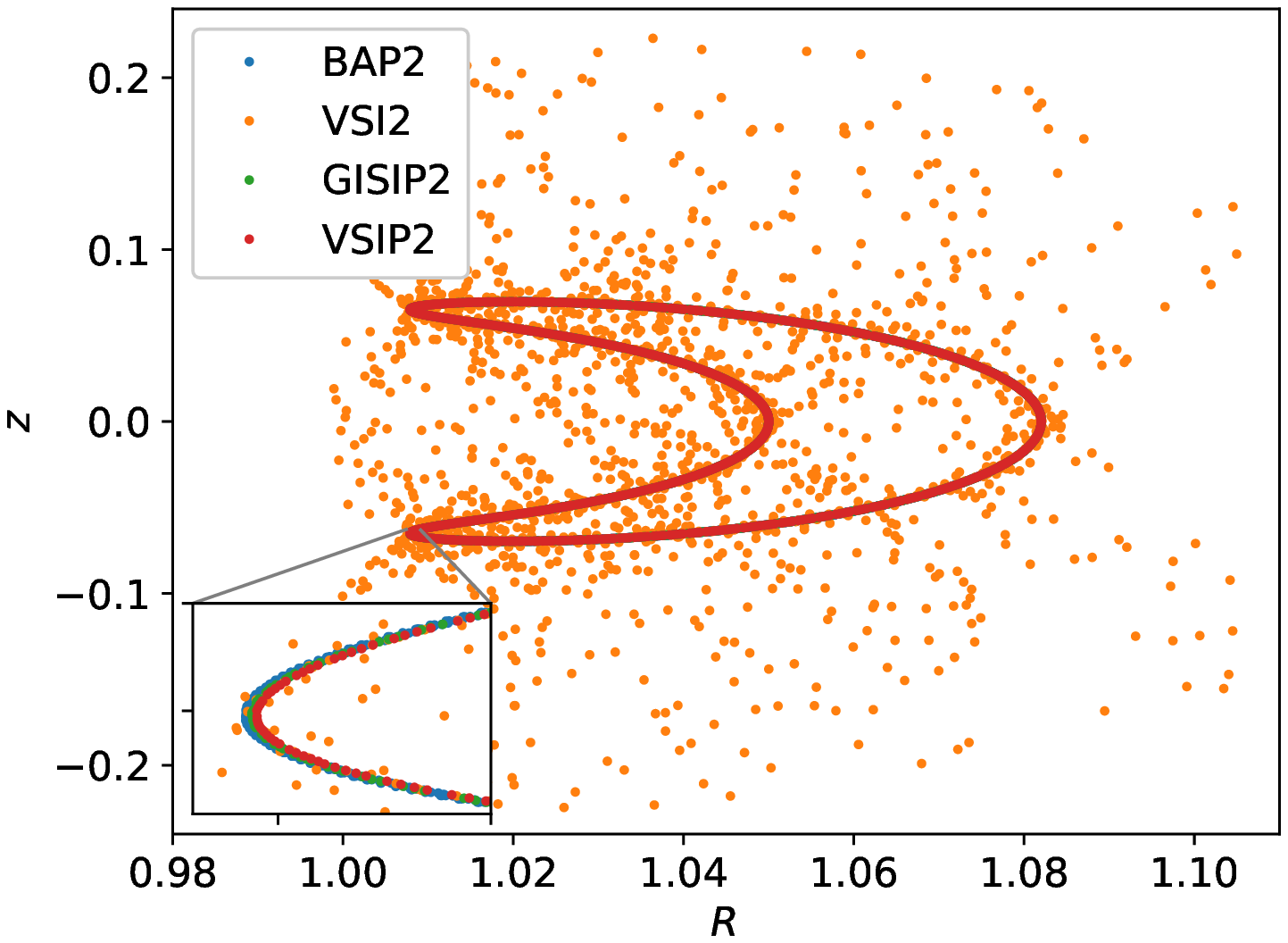}

}
\par\end{centering}
\caption{Comparison of long-term stability and accuracy of the three structure-preserving
geometric algorithms (BAP2, GISIP2 and VSIP2) for the slow manifold
dynamics of the CPP and implicit midpoint variational symplectic integrator
(VSI2) applied to the guiding center Lagrangian $L_{\mathrm{gc}}$.
The time-steps are $\Delta t=6$ (a) and $\Delta t=15$ (b), and the
total number of time-steps are $1.67\EXP{6}$ (a) and $6.67\EXP{5}$
(b). The VSI2 algorithm is unstable, while the BAP2, GISIP2, and VSIP2
algorithms are stable and accurate for long-term dynamics. }
\label{Fig6151KW} 
\end{figure}

As discussed above, the variational symplectic integrators when applied
to the guiding center Lagrangian $L_{\mathrm{gc}}$ lead to multi-step
methods due to the degeneracy of the Lagrangian, and numerical solutions
may be jeopardized by the unstable parasitic modes \citep{hairer02,hairer1999backward,Shang2011,ellison2015,lelandthesis,ellison2018}.
For the banana orbits in tokamaks, the unstable parasitic modes will
become significant when the time-step is relatively small. To demonstrate
this phenomenon, we perform long-term simulations with $\Delta t=6$
and $\Delta t=15$. The total number of time-steps is $1.67\EXP{6}$
and $6.67\EXP{5}$, respectively. The simulated orbits in $R-z$ plane
are shown in \FIG{Fig6151KW}. It can be found that orbits calculated
by the implicit midpoint variational symplectic integrator applied
to the guiding center Lagrangian $L_{\mathrm{gc}}$ (VSI2) are unstable,
while the three structure-preserving geometric algorithms for the
slow manifold dynamics of the CPP enjoy long-term stability and accuracy.

To summarize, we discovered that the slow manifolds of the classical
Pauli particle enable a family of structure-preserving geometric algorithms
for the guiding center dynamics. The mathematical difficulty associated
with the unstable parasitic modes of the discrete guiding center Lagrangian
has been overcome by the physics of the classical Pauli particle.
Unlike the degenerate guiding center Lagrangian, the classical Pauli
particle Lagrangian is regular, and variational and canonical symplectic
integrators can be directly applied without introducing unstable parasitic
modes. Three structure-preserving geometric algorithms have been implemented
for the slow manifold dynamics of the classical Pauli particle. Numerical
results confirmed that all three methods are stable with long-term
accuracy in terms of calculating slow guiding center drift motions
with time-steps significantly larger than the gyro-period. We expect
that this technique of slow manifold to be effective for other degenerate
Lagrangians reduced from regular Lagrangians. 
\begin{acknowledgments}
Jianyuan Xiao was supported by the the National MC Energy R\&D Program
(2018YFE0304100), National Key Research and Development Program (2016YFA0400600,
2016YFA0400601 and 2016YFA0400602), and the National Natural Science
Foundation of China (NSFC-11905220 and 11805273). Hong Qin was supported
by the U.S. Department of Energy (DE-AC02-09CH11466). Hong Qin thanks
Josh Burby, Lee Ellison, Alex Glasser, Yang He, Arieh Iserles, Michael
Kraus, Melvin Leok, Robert MacKay, Phil Morrison, Eric Palmerduca,
J. M. Sanz-Serna, Zaijiu Shang, Yuan Shi, Eric Sonnendr\"{u}cker,
Jonathan Squire, Yajuan Sun, Yifa Tang, Molei Tao, Ruili Zhang, and
Yao Zhou for fruitful discussions on related topics.
\end{acknowledgments}

\bibliographystyle{apsrev4-1}
\bibliography{gc2}

\begin{thebibliography}{56}%
\makeatletter
\providecommand \@ifxundefined [1]{%
 \@ifx{#1\undefined}
}%
\providecommand \@ifnum [1]{%
 \ifnum #1\expandafter \@firstoftwo
 \else \expandafter \@secondoftwo
 \fi
}%
\providecommand \@ifx [1]{%
 \ifx #1\expandafter \@firstoftwo
 \else \expandafter \@secondoftwo
 \fi
}%
\providecommand \natexlab [1]{#1}%
\providecommand \enquote  [1]{``#1''}%
\providecommand \bibnamefont  [1]{#1}%
\providecommand \bibfnamefont [1]{#1}%
\providecommand \citenamefont [1]{#1}%
\providecommand \href@noop [0]{\@secondoftwo}%
\providecommand \href [0]{\begingroup \@sanitize@url \@href}%
\providecommand \@href[1]{\@@startlink{#1}\@@href}%
\providecommand \@@href[1]{\endgroup#1\@@endlink}%
\providecommand \@sanitize@url [0]{\catcode `\\12\catcode `\$12\catcode
  `\&12\catcode `\#12\catcode `\^12\catcode `\_12\catcode `\%12\relax}%
\providecommand \@@startlink[1]{}%
\providecommand \@@endlink[0]{}%
\providecommand \url  [0]{\begingroup\@sanitize@url \@url }%
\providecommand \@url [1]{\endgroup\@href {#1}{\urlprefix }}%
\providecommand \urlprefix  [0]{URL }%
\providecommand \Eprint [0]{\href }%
\providecommand \doibase [0]{http://dx.doi.org/}%
\providecommand \selectlanguage [0]{\@gobble}%
\providecommand \bibinfo  [0]{\@secondoftwo}%
\providecommand \bibfield  [0]{\@secondoftwo}%
\providecommand \translation [1]{[#1]}%
\providecommand \BibitemOpen [0]{}%
\providecommand \bibitemStop [0]{}%
\providecommand \bibitemNoStop [0]{.\EOS\space}%
\providecommand \EOS [0]{\spacefactor3000\relax}%
\providecommand \BibitemShut  [1]{\csname bibitem#1\endcsname}%
\let\auto@bib@innerbib\@empty
\bibitem [{\citenamefont {Qin}\ and\ \citenamefont
  {Guan}(2008)}]{qin2008variational}%
  \BibitemOpen
  \bibfield  {author} {\bibinfo {author} {\bibfnamefont {H.}~\bibnamefont
  {Qin}}\ and\ \bibinfo {author} {\bibfnamefont {X.}~\bibnamefont {Guan}},\
  }\href {\doibase 10.2172/960290} {\bibfield  {journal} {\bibinfo  {journal}
  {Physical Review Letters}\ }\textbf {\bibinfo {volume} {100}},\ \bibinfo
  {pages} {035006} (\bibinfo {year} {2008})}\BibitemShut {NoStop}%
\bibitem [{\citenamefont {Qin}\ \emph {et~al.}(2009)\citenamefont {Qin},
  \citenamefont {Guan},\ and\ \citenamefont {Tang}}]{qin2009variational}%
  \BibitemOpen
  \bibfield  {author} {\bibinfo {author} {\bibfnamefont {H.}~\bibnamefont
  {Qin}}, \bibinfo {author} {\bibfnamefont {X.}~\bibnamefont {Guan}}, \ and\
  \bibinfo {author} {\bibfnamefont {W.~M.}\ \bibnamefont {Tang}},\ }\href
  {\doibase 10.1063/1.3099055} {\bibfield  {journal} {\bibinfo  {journal}
  {Physics of Plasmas (1994-present)}\ }\textbf {\bibinfo {volume} {16}},\
  \bibinfo {pages} {042510} (\bibinfo {year} {2009})}\BibitemShut {NoStop}%
\bibitem [{\citenamefont {Squire}\ \emph
  {et~al.}(2012{\natexlab{a}})\citenamefont {Squire}, \citenamefont {Qin},\
  and\ \citenamefont {Tang}}]{squire4748}%
  \BibitemOpen
  \bibfield  {author} {\bibinfo {author} {\bibfnamefont {J.}~\bibnamefont
  {Squire}}, \bibinfo {author} {\bibfnamefont {H.}~\bibnamefont {Qin}}, \ and\
  \bibinfo {author} {\bibfnamefont {W.~M.}\ \bibnamefont {Tang}},\ }\href
  {https://bp-pub.pppl.gov/pub_report//2012/PPPL-4748.pdf} {\emph {\bibinfo
  {title} {Geometric Integration of the {V}lasov-{M}axwell System with a
  Variational Particle-in-cell Scheme}}},\ \bibinfo {type} {Tech. Rep.}\
  \bibinfo {number} {PPPL-4748}\ (\bibinfo  {institution} {Princeton Plasma
  Physics Laboratory},\ \bibinfo {year} {2012})\BibitemShut {NoStop}%
\bibitem [{\citenamefont {Squire}\ \emph
  {et~al.}(2012{\natexlab{b}})\citenamefont {Squire}, \citenamefont {Qin},\
  and\ \citenamefont {Tang}}]{squire2012geometric}%
  \BibitemOpen
  \bibfield  {author} {\bibinfo {author} {\bibfnamefont {J.}~\bibnamefont
  {Squire}}, \bibinfo {author} {\bibfnamefont {H.}~\bibnamefont {Qin}}, \ and\
  \bibinfo {author} {\bibfnamefont {W.~M.}\ \bibnamefont {Tang}},\ }\href
  {\doibase 10.1063/1.4742985} {\bibfield  {journal} {\bibinfo  {journal}
  {Physics of Plasmas}\ }\textbf {\bibinfo {volume} {19}},\ \bibinfo {pages}
  {084501} (\bibinfo {year} {2012}{\natexlab{b}})}\BibitemShut {NoStop}%
\bibitem [{\citenamefont {Xiao}\ \emph {et~al.}(2013)\citenamefont {Xiao},
  \citenamefont {Liu}, \citenamefont {Qin},\ and\ \citenamefont
  {Yu}}]{xiao2013variational}%
  \BibitemOpen
  \bibfield  {author} {\bibinfo {author} {\bibfnamefont {J.}~\bibnamefont
  {Xiao}}, \bibinfo {author} {\bibfnamefont {J.}~\bibnamefont {Liu}}, \bibinfo
  {author} {\bibfnamefont {H.}~\bibnamefont {Qin}}, \ and\ \bibinfo {author}
  {\bibfnamefont {Z.}~\bibnamefont {Yu}},\ }\href {\doibase 10.1063/1.4826218}
  {\bibfield  {journal} {\bibinfo  {journal} {Physics of Plasmas}\ }\textbf
  {\bibinfo {volume} {20}},\ \bibinfo {pages} {102517} (\bibinfo {year}
  {2013})}\BibitemShut {NoStop}%
\bibitem [{\citenamefont {Xiao}\ \emph
  {et~al.}(2015{\natexlab{a}})\citenamefont {Xiao}, \citenamefont {Qin},
  \citenamefont {Liu}, \citenamefont {He}, \citenamefont {Zhang},\ and\
  \citenamefont {Sun}}]{xiao2015explicit}%
  \BibitemOpen
  \bibfield  {author} {\bibinfo {author} {\bibfnamefont {J.}~\bibnamefont
  {Xiao}}, \bibinfo {author} {\bibfnamefont {H.}~\bibnamefont {Qin}}, \bibinfo
  {author} {\bibfnamefont {J.}~\bibnamefont {Liu}}, \bibinfo {author}
  {\bibfnamefont {Y.}~\bibnamefont {He}}, \bibinfo {author} {\bibfnamefont
  {R.}~\bibnamefont {Zhang}}, \ and\ \bibinfo {author} {\bibfnamefont
  {Y.}~\bibnamefont {Sun}},\ }\href {\doibase 10.1063/1.4935904} {\bibfield
  {journal} {\bibinfo  {journal} {Physics of Plasmas}\ }\textbf {\bibinfo
  {volume} {22}},\ \bibinfo {pages} {112504} (\bibinfo {year}
  {2015}{\natexlab{a}})}\BibitemShut {NoStop}%
\bibitem [{\citenamefont {Xiao}\ \emph
  {et~al.}(2015{\natexlab{b}})\citenamefont {Xiao}, \citenamefont {Liu},
  \citenamefont {Qin}, \citenamefont {Yu},\ and\ \citenamefont
  {Xiang}}]{xiao2015variational}%
  \BibitemOpen
  \bibfield  {author} {\bibinfo {author} {\bibfnamefont {J.}~\bibnamefont
  {Xiao}}, \bibinfo {author} {\bibfnamefont {J.}~\bibnamefont {Liu}}, \bibinfo
  {author} {\bibfnamefont {H.}~\bibnamefont {Qin}}, \bibinfo {author}
  {\bibfnamefont {Z.}~\bibnamefont {Yu}}, \ and\ \bibinfo {author}
  {\bibfnamefont {N.}~\bibnamefont {Xiang}},\ }\href {\doibase
  10.1063/1.4930118} {\bibfield  {journal} {\bibinfo  {journal} {Physics of
  Plasmas}\ }\textbf {\bibinfo {volume} {22}},\ \bibinfo {pages} {092305}
  (\bibinfo {year} {2015}{\natexlab{b}})}\BibitemShut {NoStop}%
\bibitem [{\citenamefont {He}\ \emph {et~al.}(2015{\natexlab{a}})\citenamefont
  {He}, \citenamefont {Qin}, \citenamefont {Sun}, \citenamefont {Xiao},
  \citenamefont {Zhang},\ and\ \citenamefont {Liu}}]{he2015hamiltonian}%
  \BibitemOpen
  \bibfield  {author} {\bibinfo {author} {\bibfnamefont {Y.}~\bibnamefont
  {He}}, \bibinfo {author} {\bibfnamefont {H.}~\bibnamefont {Qin}}, \bibinfo
  {author} {\bibfnamefont {Y.}~\bibnamefont {Sun}}, \bibinfo {author}
  {\bibfnamefont {J.}~\bibnamefont {Xiao}}, \bibinfo {author} {\bibfnamefont
  {R.}~\bibnamefont {Zhang}}, \ and\ \bibinfo {author} {\bibfnamefont
  {J.}~\bibnamefont {Liu}},\ }\href {\doibase 10.1063/1.4938034} {\bibfield
  {journal} {\bibinfo  {journal} {Physics of Plasmas (1994-present)}\ }\textbf
  {\bibinfo {volume} {22}},\ \bibinfo {pages} {124503} (\bibinfo {year}
  {2015}{\natexlab{a}})}\BibitemShut {NoStop}%
\bibitem [{\citenamefont {Qin}\ \emph {et~al.}(2016)\citenamefont {Qin},
  \citenamefont {Liu}, \citenamefont {Xiao}, \citenamefont {Zhang},
  \citenamefont {He}, \citenamefont {Wang}, \citenamefont {Sun}, \citenamefont
  {Burby}, \citenamefont {Ellison},\ and\ \citenamefont
  {Zhou}}]{qin2016canonical}%
  \BibitemOpen
  \bibfield  {author} {\bibinfo {author} {\bibfnamefont {H.}~\bibnamefont
  {Qin}}, \bibinfo {author} {\bibfnamefont {J.}~\bibnamefont {Liu}}, \bibinfo
  {author} {\bibfnamefont {J.}~\bibnamefont {Xiao}}, \bibinfo {author}
  {\bibfnamefont {R.}~\bibnamefont {Zhang}}, \bibinfo {author} {\bibfnamefont
  {Y.}~\bibnamefont {He}}, \bibinfo {author} {\bibfnamefont {Y.}~\bibnamefont
  {Wang}}, \bibinfo {author} {\bibfnamefont {Y.}~\bibnamefont {Sun}}, \bibinfo
  {author} {\bibfnamefont {J.~W.}\ \bibnamefont {Burby}}, \bibinfo {author}
  {\bibfnamefont {L.}~\bibnamefont {Ellison}}, \ and\ \bibinfo {author}
  {\bibfnamefont {Y.}~\bibnamefont {Zhou}},\ }\href {\doibase
  10.1088/0029-5515/56/1/014001} {\bibfield  {journal} {\bibinfo  {journal}
  {Nuclear Fusion}\ }\textbf {\bibinfo {volume} {56}},\ \bibinfo {pages}
  {014001} (\bibinfo {year} {2016})}\BibitemShut {NoStop}%
\bibitem [{\citenamefont {He}\ \emph {et~al.}(2016{\natexlab{a}})\citenamefont
  {He}, \citenamefont {Sun}, \citenamefont {Qin},\ and\ \citenamefont
  {Liu}}]{he2016hamiltonian}%
  \BibitemOpen
  \bibfield  {author} {\bibinfo {author} {\bibfnamefont {Y.}~\bibnamefont
  {He}}, \bibinfo {author} {\bibfnamefont {Y.}~\bibnamefont {Sun}}, \bibinfo
  {author} {\bibfnamefont {H.}~\bibnamefont {Qin}}, \ and\ \bibinfo {author}
  {\bibfnamefont {J.}~\bibnamefont {Liu}},\ }\href {\doibase 10.1063/1.4962573}
  {\bibfield  {journal} {\bibinfo  {journal} {Physics of Plasmas}\ }\textbf
  {\bibinfo {volume} {23}},\ \bibinfo {pages} {092108} (\bibinfo {year}
  {2016}{\natexlab{a}})}\BibitemShut {NoStop}%
\bibitem [{\citenamefont {Kraus}\ \emph {et~al.}(2017)\citenamefont {Kraus},
  \citenamefont {Kormann}, \citenamefont {Morrison},\ and\ \citenamefont
  {Sonnendr\"ucker}}]{kraus2017gempic}%
  \BibitemOpen
  \bibfield  {author} {\bibinfo {author} {\bibfnamefont {M.}~\bibnamefont
  {Kraus}}, \bibinfo {author} {\bibfnamefont {K.}~\bibnamefont {Kormann}},
  \bibinfo {author} {\bibfnamefont {P.~J.}\ \bibnamefont {Morrison}}, \ and\
  \bibinfo {author} {\bibfnamefont {E.}~\bibnamefont {Sonnendr\"ucker}},\
  }\href {\doibase 10.1017/s002237781700040x} {\bibfield  {journal} {\bibinfo
  {journal} {Journal of Plasma Physics}\ }\textbf {\bibinfo {volume} {83}},\
  \bibinfo {pages} {905830401} (\bibinfo {year} {2017})}\BibitemShut {NoStop}%
\bibitem [{\citenamefont {Xiao}\ \emph {et~al.}(2017)\citenamefont {Xiao},
  \citenamefont {Qin}, \citenamefont {Liu},\ and\ \citenamefont
  {Zhang}}]{xiao2017local}%
  \BibitemOpen
  \bibfield  {author} {\bibinfo {author} {\bibfnamefont {J.}~\bibnamefont
  {Xiao}}, \bibinfo {author} {\bibfnamefont {H.}~\bibnamefont {Qin}}, \bibinfo
  {author} {\bibfnamefont {J.}~\bibnamefont {Liu}}, \ and\ \bibinfo {author}
  {\bibfnamefont {R.}~\bibnamefont {Zhang}},\ }\href {\doibase
  10.1063/1.4986097} {\bibfield  {journal} {\bibinfo  {journal} {Physics of
  Plasmas}\ }\textbf {\bibinfo {volume} {24}},\ \bibinfo {pages} {062112}
  (\bibinfo {year} {2017})}\BibitemShut {NoStop}%
\bibitem [{\citenamefont {Xiao}\ \emph {et~al.}(2018)\citenamefont {Xiao},
  \citenamefont {Qin},\ and\ \citenamefont {Liu}}]{xiao2018structure}%
  \BibitemOpen
  \bibfield  {author} {\bibinfo {author} {\bibfnamefont {J.}~\bibnamefont
  {Xiao}}, \bibinfo {author} {\bibfnamefont {H.}~\bibnamefont {Qin}}, \ and\
  \bibinfo {author} {\bibfnamefont {J.}~\bibnamefont {Liu}},\ }\href {\doibase
  10.1088/2058-6272/aac3d1} {\bibfield  {journal} {\bibinfo  {journal} {Plasma
  Science and Technology}\ }\textbf {\bibinfo {volume} {20}},\ \bibinfo {eid}
  {110501} (\bibinfo {year} {2018})}\BibitemShut {NoStop}%
\bibitem [{\citenamefont {Xiao}\ and\ \citenamefont
  {Qin}(2019{\natexlab{a}})}]{xiao2019field}%
  \BibitemOpen
  \bibfield  {author} {\bibinfo {author} {\bibfnamefont {J.}~\bibnamefont
  {Xiao}}\ and\ \bibinfo {author} {\bibfnamefont {H.}~\bibnamefont {Qin}},\
  }\href {\doibase 10.1088/1741-4326/ab38dc} {\bibfield  {journal} {\bibinfo
  {journal} {Nuclear Fusion}\ }\textbf {\bibinfo {volume} {59}},\ \bibinfo
  {pages} {106044} (\bibinfo {year} {2019}{\natexlab{a}})}\BibitemShut
  {NoStop}%
\bibitem [{\citenamefont {Xiao}\ and\ \citenamefont {Qin}(2020)}]{Xiao2020}%
  \BibitemOpen
  \bibfield  {author} {\bibinfo {author} {\bibfnamefont {J.}~\bibnamefont
  {Xiao}}\ and\ \bibinfo {author} {\bibfnamefont {H.}~\bibnamefont {Qin}},\
  }\href@noop {} {\enquote {\bibinfo {title} {Explicit structure-preserving
  geometric particle-in-cell algorithm in curvilinear orthogonal coordinate
  systems and its applications to whole-device 6{D} kinetic simulations of
  tokamak physics},}\ } (\bibinfo {year} {2020}),\ \Eprint
  {http://arxiv.org/abs/2004.08150v1} {2004.08150v1} \BibitemShut {NoStop}%
\bibitem [{\citenamefont {Glasser}\ and\ \citenamefont
  {Qin}(2020)}]{Glasser2020}%
  \BibitemOpen
  \bibfield  {author} {\bibinfo {author} {\bibfnamefont {A.~S.}\ \bibnamefont
  {Glasser}}\ and\ \bibinfo {author} {\bibfnamefont {H.}~\bibnamefont {Qin}},\
  }\href {\doibase 10.1017/s0022377820000434} {\bibfield  {journal} {\bibinfo
  {journal} {Journal of Plasma Physics}\ }\textbf {\bibinfo {volume} {86}},\
  \bibinfo {pages} {835860303} (\bibinfo {year} {2020})}\BibitemShut {NoStop}%
\bibitem [{\citenamefont {Mouhot}\ and\ \citenamefont
  {Villani}(2011)}]{Mouhot2011}%
  \BibitemOpen
  \bibfield  {author} {\bibinfo {author} {\bibfnamefont {C.}~\bibnamefont
  {Mouhot}}\ and\ \bibinfo {author} {\bibfnamefont {C.}~\bibnamefont
  {Villani}},\ }\href {\doibase 10.1007/s11511-011-0068-9} {\bibfield
  {journal} {\bibinfo  {journal} {Acta Mathematica}\ }\textbf {\bibinfo
  {volume} {207}},\ \bibinfo {pages} {29} (\bibinfo {year} {2011})}\BibitemShut
  {NoStop}%
\bibitem [{\citenamefont {Shi}\ \emph {et~al.}(2018)\citenamefont {Shi},
  \citenamefont {Xiao}, \citenamefont {Qin},\ and\ \citenamefont
  {Fisch}}]{shi2018}%
  \BibitemOpen
  \bibfield  {author} {\bibinfo {author} {\bibfnamefont {Y.}~\bibnamefont
  {Shi}}, \bibinfo {author} {\bibfnamefont {J.}~\bibnamefont {Xiao}}, \bibinfo
  {author} {\bibfnamefont {H.}~\bibnamefont {Qin}}, \ and\ \bibinfo {author}
  {\bibfnamefont {N.~J.}\ \bibnamefont {Fisch}},\ }\href {\doibase
  10.1103/physreve.97.053206} {\bibfield  {journal} {\bibinfo  {journal}
  {Physical Review E}\ }\textbf {\bibinfo {volume} {97}},\ \bibinfo {pages}
  {053206} (\bibinfo {year} {2018})}\BibitemShut {NoStop}%
\bibitem [{\citenamefont {Zhou}\ \emph
  {et~al.}(2017{\natexlab{a}})\citenamefont {Zhou}, \citenamefont {Huang},
  \citenamefont {Qin},\ and\ \citenamefont {Bhattacharjee}}]{Zhou2017APJ}%
  \BibitemOpen
  \bibfield  {author} {\bibinfo {author} {\bibfnamefont {Y.}~\bibnamefont
  {Zhou}}, \bibinfo {author} {\bibfnamefont {Y.-M.}\ \bibnamefont {Huang}},
  \bibinfo {author} {\bibfnamefont {H.}~\bibnamefont {Qin}}, \ and\ \bibinfo
  {author} {\bibfnamefont {A.}~\bibnamefont {Bhattacharjee}},\ }\href {\doibase
  10.3847/1538-4357/aa9b84} {\bibfield  {journal} {\bibinfo  {journal} {The
  Astrophysical Journal}\ }\textbf {\bibinfo {volume} {852}},\ \bibinfo {pages}
  {3} (\bibinfo {year} {2017}{\natexlab{a}})}\BibitemShut {NoStop}%
\bibitem [{\citenamefont {Ruth}(1983)}]{ruth83}%
  \BibitemOpen
  \bibfield  {author} {\bibinfo {author} {\bibfnamefont {R.~D.}\ \bibnamefont
  {Ruth}},\ }\href {\doibase 10.1109/tns.1983.4332919} {\bibfield  {journal}
  {\bibinfo  {journal} {IEEE Trans. Nucl. Sci}\ }\textbf {\bibinfo {volume}
  {30}},\ \bibinfo {pages} {2669} (\bibinfo {year} {1983})}\BibitemShut
  {NoStop}%
\bibitem [{\citenamefont {Feng}(1985)}]{feng85}%
  \BibitemOpen
  \bibfield  {author} {\bibinfo {author} {\bibfnamefont {K.}~\bibnamefont
  {Feng}},\ }in\ \href@noop {} {\emph {\bibinfo {booktitle} {the Proceedings of
  1984 Beijing Symposium on Differential Geometry and Differential
  Equations}}},\ \bibinfo {editor} {edited by\ \bibinfo {editor} {\bibfnamefont
  {K.}~\bibnamefont {Feng}}}\ (\bibinfo  {publisher} {Science Press},\ \bibinfo
  {year} {1985})\ pp.\ \bibinfo {pages} {42--58}\BibitemShut {NoStop}%
\bibitem [{\citenamefont {Feng}(1986)}]{feng86}%
  \BibitemOpen
  \bibfield  {author} {\bibinfo {author} {\bibfnamefont {K.}~\bibnamefont
  {Feng}},\ }\href@noop {} {\bibfield  {journal} {\bibinfo  {journal} {J.
  Comput. Maths.}\ }\textbf {\bibinfo {volume} {4}},\ \bibinfo {pages} {279}
  (\bibinfo {year} {1986})}\BibitemShut {NoStop}%
\bibitem [{\citenamefont {Sanz-Serna}(1988)}]{SanzSerna1988}%
  \BibitemOpen
  \bibfield  {author} {\bibinfo {author} {\bibfnamefont {J.~M.}\ \bibnamefont
  {Sanz-Serna}},\ }\href {\doibase 10.1007/bf01954907} {\bibfield  {journal}
  {\bibinfo  {journal} {{BIT}}\ }\textbf {\bibinfo {volume} {28}},\ \bibinfo
  {pages} {877} (\bibinfo {year} {1988})}\BibitemShut {NoStop}%
\bibitem [{\citenamefont {Feng}\ and\ \citenamefont {Qin}(2010)}]{feng10}%
  \BibitemOpen
  \bibfield  {author} {\bibinfo {author} {\bibfnamefont {K.}~\bibnamefont
  {Feng}}\ and\ \bibinfo {author} {\bibfnamefont {M.}~\bibnamefont {Qin}},\
  }\href@noop {} {\emph {\bibinfo {title} {Symplectic Geometric Algorithms for
  Hamiltonian Systems}}}\ (\bibinfo  {publisher} {Springer-Verlag},\ \bibinfo
  {year} {2010})\BibitemShut {NoStop}%
\bibitem [{\citenamefont {Sanz-Serna}\ and\ \citenamefont
  {Calvo}(1994)}]{sanz-serna94}%
  \BibitemOpen
  \bibfield  {author} {\bibinfo {author} {\bibfnamefont {J.~M.}\ \bibnamefont
  {Sanz-Serna}}\ and\ \bibinfo {author} {\bibfnamefont {M.~P.}\ \bibnamefont
  {Calvo}},\ }\href@noop {} {\emph {\bibinfo {title} {Numerical Hamiltonian
  Problems}}}\ (\bibinfo  {publisher} {Chapman and Hall},\ \bibinfo {address}
  {London},\ \bibinfo {year} {1994})\BibitemShut {NoStop}%
\bibitem [{\citenamefont {Hairer}\ \emph {et~al.}(2002)\citenamefont {Hairer},
  \citenamefont {Lubich},\ and\ \citenamefont {Wanner}}]{hairer02}%
  \BibitemOpen
  \bibfield  {author} {\bibinfo {author} {\bibfnamefont {E.}~\bibnamefont
  {Hairer}}, \bibinfo {author} {\bibfnamefont {C.}~\bibnamefont {Lubich}}, \
  and\ \bibinfo {author} {\bibfnamefont {G.}~\bibnamefont {Wanner}},\
  }\href@noop {} {\emph {\bibinfo {title} {Geometric Numerical Integration:
  Structure-Preserving Algorithms for Ordinary Differential Equations}}}\
  (\bibinfo  {publisher} {Springer},\ \bibinfo {address} {New York},\ \bibinfo
  {year} {2002})\ pp.\ \bibinfo {pages} {567--616}\BibitemShut {NoStop}%
\bibitem [{\citenamefont {Marsden}\ and\ \citenamefont
  {West}(2001)}]{marsden2001discrete}%
  \BibitemOpen
  \bibfield  {author} {\bibinfo {author} {\bibfnamefont {J.~E.}\ \bibnamefont
  {Marsden}}\ and\ \bibinfo {author} {\bibfnamefont {M.}~\bibnamefont {West}},\
  }\href {\doibase 10.1017/S096249290100006X} {\bibfield  {journal} {\bibinfo
  {journal} {Acta Numer.}\ }\textbf {\bibinfo {volume} {10}},\ \bibinfo {pages}
  {357} (\bibinfo {year} {2001})}\BibitemShut {NoStop}%
\bibitem [{\citenamefont {Shang}(2011)}]{Shang2011}%
  \BibitemOpen
  \bibfield  {author} {\bibinfo {author} {\bibfnamefont {Z.}~\bibnamefont
  {Shang}},\ }\href@noop {} {} (\bibinfo {year} {2011}),\ \bibinfo {note}
  {private communication}\BibitemShut {NoStop}%
\bibitem [{\citenamefont {Ellison}\ \emph {et~al.}(2015)\citenamefont
  {Ellison}, \citenamefont {Finn}, \citenamefont {Qin},\ and\ \citenamefont
  {Tang}}]{ellison2015}%
  \BibitemOpen
  \bibfield  {author} {\bibinfo {author} {\bibfnamefont {C.~L.}\ \bibnamefont
  {Ellison}}, \bibinfo {author} {\bibfnamefont {J.~M.}\ \bibnamefont {Finn}},
  \bibinfo {author} {\bibfnamefont {H.}~\bibnamefont {Qin}}, \ and\ \bibinfo
  {author} {\bibfnamefont {W.~M.}\ \bibnamefont {Tang}},\ }\href {\doibase
  10.1088/0741-3335/57/5/054007} {\bibfield  {journal} {\bibinfo  {journal}
  {Plasma Physics and Controlled Fusion}\ }\textbf {\bibinfo {volume} {57}},\
  \bibinfo {pages} {054007} (\bibinfo {year} {2015})}\BibitemShut {NoStop}%
\bibitem [{\citenamefont {Ellison}(2016)}]{lelandthesis}%
  \BibitemOpen
  \bibfield  {author} {\bibinfo {author} {\bibfnamefont {C.~L.}\ \bibnamefont
  {Ellison}},\ }\emph {\bibinfo {title} {Development of Multistep and
  Degenerate Variational Integrators for Applications in Plasma Physics}},\
  \href@noop {} {Ph.D. thesis},\ \bibinfo  {school} {Princeton University}
  (\bibinfo {year} {2016})\BibitemShut {NoStop}%
\bibitem [{\citenamefont {Zhang}\ \emph {et~al.}(2014)\citenamefont {Zhang},
  \citenamefont {Liu}, \citenamefont {Tang}, \citenamefont {Qin}, \citenamefont
  {Xiao},\ and\ \citenamefont {Zhu}}]{zhang2014canonicalization}%
  \BibitemOpen
  \bibfield  {author} {\bibinfo {author} {\bibfnamefont {R.}~\bibnamefont
  {Zhang}}, \bibinfo {author} {\bibfnamefont {J.}~\bibnamefont {Liu}}, \bibinfo
  {author} {\bibfnamefont {Y.}~\bibnamefont {Tang}}, \bibinfo {author}
  {\bibfnamefont {H.}~\bibnamefont {Qin}}, \bibinfo {author} {\bibfnamefont
  {J.}~\bibnamefont {Xiao}}, \ and\ \bibinfo {author} {\bibfnamefont
  {B.}~\bibnamefont {Zhu}},\ }\href {\doibase 10.1063/1.4867669} {\bibfield
  {journal} {\bibinfo  {journal} {Physics of Plasmas (1994-present)}\ }\textbf
  {\bibinfo {volume} {21}},\ \bibinfo {pages} {032504} (\bibinfo {year}
  {2014})}\BibitemShut {NoStop}%
\bibitem [{\citenamefont {Burby}\ and\ \citenamefont
  {Ellison}(2017)}]{burby2017}%
  \BibitemOpen
  \bibfield  {author} {\bibinfo {author} {\bibfnamefont {J.}~\bibnamefont
  {Burby}}\ and\ \bibinfo {author} {\bibfnamefont {C.}~\bibnamefont
  {Ellison}},\ }\href {\doibase 10.1063/1.5004429} {\bibfield  {journal}
  {\bibinfo  {journal} {Physics of Plasmas}\ }\textbf {\bibinfo {volume}
  {24}},\ \bibinfo {pages} {110703} (\bibinfo {year} {2017})}\BibitemShut
  {NoStop}%
\bibitem [{\citenamefont {Kraus}(2017)}]{kraus2017}%
  \BibitemOpen
  \bibfield  {author} {\bibinfo {author} {\bibfnamefont {M.}~\bibnamefont
  {Kraus}},\ }\href@noop {} {\enquote {\bibinfo {title} {Projected variational
  integrators for degenerate {L}agrangian systems},}\ } (\bibinfo {year}
  {2017}),\ \Eprint {http://arxiv.org/abs/1708.07356v1} {arXiv:1708.07356v1}
  \BibitemShut {NoStop}%
\bibitem [{\citenamefont {Ellison}\ \emph {et~al.}(2018)\citenamefont
  {Ellison}, \citenamefont {Finn}, \citenamefont {Burby}, \citenamefont
  {Kraus}, \citenamefont {Qin},\ and\ \citenamefont {Tang}}]{ellison2018}%
  \BibitemOpen
  \bibfield  {author} {\bibinfo {author} {\bibfnamefont {C.~L.}\ \bibnamefont
  {Ellison}}, \bibinfo {author} {\bibfnamefont {J.~M.}\ \bibnamefont {Finn}},
  \bibinfo {author} {\bibfnamefont {J.~W.}\ \bibnamefont {Burby}}, \bibinfo
  {author} {\bibfnamefont {M.}~\bibnamefont {Kraus}}, \bibinfo {author}
  {\bibfnamefont {H.}~\bibnamefont {Qin}}, \ and\ \bibinfo {author}
  {\bibfnamefont {W.~M.}\ \bibnamefont {Tang}},\ }\href {\doibase
  10.1063/1.5022277} {\bibfield  {journal} {\bibinfo  {journal} {Physics of
  Plasmas}\ }\textbf {\bibinfo {volume} {25}},\ \bibinfo {pages} {052502}
  (\bibinfo {year} {2018})}\BibitemShut {NoStop}%
\bibitem [{\citenamefont {Qin}\ \emph {et~al.}(2013)\citenamefont {Qin},
  \citenamefont {Zhang}, \citenamefont {Xiao}, \citenamefont {Liu},
  \citenamefont {Sun},\ and\ \citenamefont {Tang}}]{qin2013boris}%
  \BibitemOpen
  \bibfield  {author} {\bibinfo {author} {\bibfnamefont {H.}~\bibnamefont
  {Qin}}, \bibinfo {author} {\bibfnamefont {S.}~\bibnamefont {Zhang}}, \bibinfo
  {author} {\bibfnamefont {J.}~\bibnamefont {Xiao}}, \bibinfo {author}
  {\bibfnamefont {J.}~\bibnamefont {Liu}}, \bibinfo {author} {\bibfnamefont
  {Y.}~\bibnamefont {Sun}}, \ and\ \bibinfo {author} {\bibfnamefont {W.~M.}\
  \bibnamefont {Tang}},\ }\href {\doibase 10.2172/1090047} {\bibfield
  {journal} {\bibinfo  {journal} {Physics of Plasmas (1994-present)}\ }\textbf
  {\bibinfo {volume} {20}},\ \bibinfo {pages} {084503} (\bibinfo {year}
  {2013})}\BibitemShut {NoStop}%
\bibitem [{\citenamefont {Zhang}\ \emph {et~al.}(2015)\citenamefont {Zhang},
  \citenamefont {Liu}, \citenamefont {Qin}, \citenamefont {Wang}, \citenamefont
  {He},\ and\ \citenamefont {Sun}}]{zhang2015volume}%
  \BibitemOpen
  \bibfield  {author} {\bibinfo {author} {\bibfnamefont {R.}~\bibnamefont
  {Zhang}}, \bibinfo {author} {\bibfnamefont {J.}~\bibnamefont {Liu}}, \bibinfo
  {author} {\bibfnamefont {H.}~\bibnamefont {Qin}}, \bibinfo {author}
  {\bibfnamefont {Y.}~\bibnamefont {Wang}}, \bibinfo {author} {\bibfnamefont
  {Y.}~\bibnamefont {He}}, \ and\ \bibinfo {author} {\bibfnamefont
  {Y.}~\bibnamefont {Sun}},\ }\href {\doibase 10.1063/1.4916570} {\bibfield
  {journal} {\bibinfo  {journal} {Physics of Plasmas (1994-present)}\ }\textbf
  {\bibinfo {volume} {22}},\ \bibinfo {pages} {044501} (\bibinfo {year}
  {2015})}\BibitemShut {NoStop}%
\bibitem [{\citenamefont {He}\ \emph {et~al.}(2015{\natexlab{b}})\citenamefont
  {He}, \citenamefont {Sun}, \citenamefont {Liu},\ and\ \citenamefont
  {Qin}}]{he2015volume}%
  \BibitemOpen
  \bibfield  {author} {\bibinfo {author} {\bibfnamefont {Y.}~\bibnamefont
  {He}}, \bibinfo {author} {\bibfnamefont {Y.}~\bibnamefont {Sun}}, \bibinfo
  {author} {\bibfnamefont {J.}~\bibnamefont {Liu}}, \ and\ \bibinfo {author}
  {\bibfnamefont {H.}~\bibnamefont {Qin}},\ }\href {\doibase
  10.1016/j.jcp.2014.10.032} {\bibfield  {journal} {\bibinfo  {journal}
  {Journal of Computational Physics}\ }\textbf {\bibinfo {volume} {281}},\
  \bibinfo {pages} {135} (\bibinfo {year} {2015}{\natexlab{b}})}\BibitemShut
  {NoStop}%
\bibitem [{\citenamefont {He}\ \emph {et~al.}(2016{\natexlab{b}})\citenamefont
  {He}, \citenamefont {Sun}, \citenamefont {Zhang}, \citenamefont {Wang},
  \citenamefont {Liu},\ and\ \citenamefont {Qin}}]{he2016high}%
  \BibitemOpen
  \bibfield  {author} {\bibinfo {author} {\bibfnamefont {Y.}~\bibnamefont
  {He}}, \bibinfo {author} {\bibfnamefont {Y.}~\bibnamefont {Sun}}, \bibinfo
  {author} {\bibfnamefont {R.}~\bibnamefont {Zhang}}, \bibinfo {author}
  {\bibfnamefont {Y.}~\bibnamefont {Wang}}, \bibinfo {author} {\bibfnamefont
  {J.}~\bibnamefont {Liu}}, \ and\ \bibinfo {author} {\bibfnamefont
  {H.}~\bibnamefont {Qin}},\ }\href {\doibase 10.1063/1.4962677} {\bibfield
  {journal} {\bibinfo  {journal} {Physics of Plasmas}\ }\textbf {\bibinfo
  {volume} {23}},\ \bibinfo {pages} {092109} (\bibinfo {year}
  {2016}{\natexlab{b}})}\BibitemShut {NoStop}%
\bibitem [{\citenamefont {He}\ \emph {et~al.}(2016{\natexlab{c}})\citenamefont
  {He}, \citenamefont {Sun}, \citenamefont {Liu},\ and\ \citenamefont
  {Qin}}]{he2016higher}%
  \BibitemOpen
  \bibfield  {author} {\bibinfo {author} {\bibfnamefont {Y.}~\bibnamefont
  {He}}, \bibinfo {author} {\bibfnamefont {Y.}~\bibnamefont {Sun}}, \bibinfo
  {author} {\bibfnamefont {J.}~\bibnamefont {Liu}}, \ and\ \bibinfo {author}
  {\bibfnamefont {H.}~\bibnamefont {Qin}},\ }\href {\doibase
  10.1016/j.jcp.2015.10.032} {\bibfield  {journal} {\bibinfo  {journal}
  {Journal of Computational Physics}\ }\textbf {\bibinfo {volume} {305}},\
  \bibinfo {pages} {172 } (\bibinfo {year} {2016}{\natexlab{c}})}\BibitemShut
  {NoStop}%
\bibitem [{\citenamefont {Zhang}\ \emph {et~al.}(2016)\citenamefont {Zhang},
  \citenamefont {Qin}, \citenamefont {Tang}, \citenamefont {Liu}, \citenamefont
  {He},\ and\ \citenamefont {Xiao}}]{zhang2016explicit}%
  \BibitemOpen
  \bibfield  {author} {\bibinfo {author} {\bibfnamefont {R.}~\bibnamefont
  {Zhang}}, \bibinfo {author} {\bibfnamefont {H.}~\bibnamefont {Qin}}, \bibinfo
  {author} {\bibfnamefont {Y.}~\bibnamefont {Tang}}, \bibinfo {author}
  {\bibfnamefont {J.}~\bibnamefont {Liu}}, \bibinfo {author} {\bibfnamefont
  {Y.}~\bibnamefont {He}}, \ and\ \bibinfo {author} {\bibfnamefont
  {J.}~\bibnamefont {Xiao}},\ }\href {\doibase 10.1103/physreve.94.013205}
  {\bibfield  {journal} {\bibinfo  {journal} {Physical Review E}\ }\textbf
  {\bibinfo {volume} {94}},\ \bibinfo {pages} {013205} (\bibinfo {year}
  {2016})}\BibitemShut {NoStop}%
\bibitem [{\citenamefont {Tu}\ \emph {et~al.}(2016)\citenamefont {Tu},
  \citenamefont {Zhu}, \citenamefont {Tang}, \citenamefont {Qin}, \citenamefont
  {Liu},\ and\ \citenamefont {Zhang}}]{tu2016}%
  \BibitemOpen
  \bibfield  {author} {\bibinfo {author} {\bibfnamefont {X.}~\bibnamefont
  {Tu}}, \bibinfo {author} {\bibfnamefont {B.}~\bibnamefont {Zhu}}, \bibinfo
  {author} {\bibfnamefont {Y.}~\bibnamefont {Tang}}, \bibinfo {author}
  {\bibfnamefont {H.}~\bibnamefont {Qin}}, \bibinfo {author} {\bibfnamefont
  {J.}~\bibnamefont {Liu}}, \ and\ \bibinfo {author} {\bibfnamefont
  {R.}~\bibnamefont {Zhang}},\ }\href {\doibase 10.1063/1.4972878} {\bibfield
  {journal} {\bibinfo  {journal} {Physics of Plasmas}\ }\textbf {\bibinfo
  {volume} {23}},\ \bibinfo {pages} {122514} (\bibinfo {year}
  {2016})}\BibitemShut {NoStop}%
\bibitem [{\citenamefont {Tao}(2016)}]{tao2016}%
  \BibitemOpen
  \bibfield  {author} {\bibinfo {author} {\bibfnamefont {M.}~\bibnamefont
  {Tao}},\ }\href {\doibase 10.1016/j.jcp.2016.09.047} {\bibfield  {journal}
  {\bibinfo  {journal} {Journal of Computational Physics}\ }\textbf {\bibinfo
  {volume} {327}},\ \bibinfo {pages} {245} (\bibinfo {year}
  {2016})}\BibitemShut {NoStop}%
\bibitem [{\citenamefont {He}\ \emph {et~al.}(2017)\citenamefont {He},
  \citenamefont {Zhou}, \citenamefont {Sun}, \citenamefont {Liu},\ and\
  \citenamefont {Qin}}]{he2017explicit}%
  \BibitemOpen
  \bibfield  {author} {\bibinfo {author} {\bibfnamefont {Y.}~\bibnamefont
  {He}}, \bibinfo {author} {\bibfnamefont {Z.}~\bibnamefont {Zhou}}, \bibinfo
  {author} {\bibfnamefont {Y.}~\bibnamefont {Sun}}, \bibinfo {author}
  {\bibfnamefont {J.}~\bibnamefont {Liu}}, \ and\ \bibinfo {author}
  {\bibfnamefont {H.}~\bibnamefont {Qin}},\ }\href {\doibase
  10.1016/j.physleta.2016.12.031} {\bibfield  {journal} {\bibinfo  {journal}
  {Physics Letters A}\ }\textbf {\bibinfo {volume} {381}},\ \bibinfo {pages}
  {568} (\bibinfo {year} {2017})}\BibitemShut {NoStop}%
\bibitem [{\citenamefont {Zhou}\ \emph
  {et~al.}(2017{\natexlab{b}})\citenamefont {Zhou}, \citenamefont {He},
  \citenamefont {Sun}, \citenamefont {Liu},\ and\ \citenamefont
  {Qin}}]{zhou2017explicit}%
  \BibitemOpen
  \bibfield  {author} {\bibinfo {author} {\bibfnamefont {Z.}~\bibnamefont
  {Zhou}}, \bibinfo {author} {\bibfnamefont {Y.}~\bibnamefont {He}}, \bibinfo
  {author} {\bibfnamefont {Y.}~\bibnamefont {Sun}}, \bibinfo {author}
  {\bibfnamefont {J.}~\bibnamefont {Liu}}, \ and\ \bibinfo {author}
  {\bibfnamefont {H.}~\bibnamefont {Qin}},\ }\href {\doibase 10.1063/1.4982743}
  {\bibfield  {journal} {\bibinfo  {journal} {Physics of Plasmas}\ }\textbf
  {\bibinfo {volume} {24}},\ \bibinfo {pages} {052507} (\bibinfo {year}
  {2017}{\natexlab{b}})}\BibitemShut {NoStop}%
\bibitem [{\citenamefont {Xiao}\ and\ \citenamefont
  {Qin}(2019{\natexlab{b}})}]{xiao2019a}%
  \BibitemOpen
  \bibfield  {author} {\bibinfo {author} {\bibfnamefont {J.}~\bibnamefont
  {Xiao}}\ and\ \bibinfo {author} {\bibfnamefont {H.}~\bibnamefont {Qin}},\
  }\href {\doibase 10.1016/j.cpc.2019.04.003} {\bibfield  {journal} {\bibinfo
  {journal} {Computer Physics Communications}\ }\textbf {\bibinfo {volume}
  {241}},\ \bibinfo {pages} {19} (\bibinfo {year}
  {2019}{\natexlab{b}})}\BibitemShut {NoStop}%
\bibitem [{\citenamefont {Shi}\ \emph {et~al.}(2019)\citenamefont {Shi},
  \citenamefont {Sun}, \citenamefont {He}, \citenamefont {Qin},\ and\
  \citenamefont {Liu}}]{shi2019}%
  \BibitemOpen
  \bibfield  {author} {\bibinfo {author} {\bibfnamefont {Y.}~\bibnamefont
  {Shi}}, \bibinfo {author} {\bibfnamefont {Y.}~\bibnamefont {Sun}}, \bibinfo
  {author} {\bibfnamefont {Y.}~\bibnamefont {He}}, \bibinfo {author}
  {\bibfnamefont {H.}~\bibnamefont {Qin}}, \ and\ \bibinfo {author}
  {\bibfnamefont {J.}~\bibnamefont {Liu}},\ }\href {\doibase
  10.1007/s11075-018-0636-6} {\bibfield  {journal} {\bibinfo  {journal}
  {Numerical Algorithms}\ }\textbf {\bibinfo {volume} {81}},\ \bibinfo {pages}
  {1295} (\bibinfo {year} {2019})}\BibitemShut {NoStop}%
\bibitem [{\citenamefont {Littlejohn}(1983)}]{littlejohn1983variational}%
  \BibitemOpen
  \bibfield  {author} {\bibinfo {author} {\bibfnamefont {R.~G.}\ \bibnamefont
  {Littlejohn}},\ }\href {\doibase 10.1017/S002237780000060X} {\bibfield
  {journal} {\bibinfo  {journal} {Journal of Plasma Physics}\ }\textbf
  {\bibinfo {volume} {29}},\ \bibinfo {pages} {111} (\bibinfo {year}
  {1983})}\BibitemShut {NoStop}%
\bibitem [{\citenamefont {Qin}\ and\ \citenamefont {Davidson}(2006)}]{Qin2006}%
  \BibitemOpen
  \bibfield  {author} {\bibinfo {author} {\bibfnamefont {H.}~\bibnamefont
  {Qin}}\ and\ \bibinfo {author} {\bibfnamefont {R.~C.}\ \bibnamefont
  {Davidson}},\ }\href {\doibase 10.1103/physrevlett.96.085003} {\bibfield
  {journal} {\bibinfo  {journal} {Physical Review Letters}\ }\textbf {\bibinfo
  {volume} {96}},\ \bibinfo {pages} {085003} (\bibinfo {year}
  {2006})}\BibitemShut {NoStop}%
\bibitem [{\citenamefont {Lorenz}(1986)}]{Lorenz_1986}%
  \BibitemOpen
  \bibfield  {author} {\bibinfo {author} {\bibfnamefont {E.~N.}\ \bibnamefont
  {Lorenz}},\ }\href {\doibase 10.1175/1520-0469(1986)043<1547:oteoas>2.0.co;2}
  {\bibfield  {journal} {\bibinfo  {journal} {Journal of the Atmospheric
  Sciences}\ }\textbf {\bibinfo {volume} {43}},\ \bibinfo {pages} {1547}
  (\bibinfo {year} {1986})}\BibitemShut {NoStop}%
\bibitem [{\citenamefont {MacKay}(2004)}]{MacKay2004}%
  \BibitemOpen
  \bibfield  {author} {\bibinfo {author} {\bibfnamefont {R.~S.}\ \bibnamefont
  {MacKay}},\ }in\ \href {\doibase 10.1142/9789812794864_0003} {\emph {\bibinfo
  {booktitle} {Energy Localisation and Transfer}}}\ (\bibinfo  {publisher}
  {World Scientific},\ \bibinfo {year} {2004})\ pp.\ \bibinfo {pages}
  {149--192}\BibitemShut {NoStop}%
\bibitem [{\citenamefont {Burby}(2020)}]{Burby2020}%
  \BibitemOpen
  \bibfield  {author} {\bibinfo {author} {\bibfnamefont {J.~W.}\ \bibnamefont
  {Burby}},\ }\href {\doibase 10.1063/1.5119801} {\bibfield  {journal}
  {\bibinfo  {journal} {Journal of Mathematical Physics}\ }\textbf {\bibinfo
  {volume} {61}},\ \bibinfo {pages} {012703} (\bibinfo {year}
  {2020})}\BibitemShut {NoStop}%
\bibitem [{\citenamefont {Li}\ \emph {et~al.}(2011)\citenamefont {Li},
  \citenamefont {Qin}, \citenamefont {Pu}, \citenamefont {Xie},\ and\
  \citenamefont {Fu}}]{li2011variational}%
  \BibitemOpen
  \bibfield  {author} {\bibinfo {author} {\bibfnamefont {J.}~\bibnamefont
  {Li}}, \bibinfo {author} {\bibfnamefont {H.}~\bibnamefont {Qin}}, \bibinfo
  {author} {\bibfnamefont {Z.}~\bibnamefont {Pu}}, \bibinfo {author}
  {\bibfnamefont {L.}~\bibnamefont {Xie}}, \ and\ \bibinfo {author}
  {\bibfnamefont {S.}~\bibnamefont {Fu}},\ }\href {\doibase 10.1063/1.3589275}
  {\bibfield  {journal} {\bibinfo  {journal} {Physics of Plasmas}\ }\textbf
  {\bibinfo {volume} {18}},\ \bibinfo {pages} {052902} (\bibinfo {year}
  {2011})}\BibitemShut {NoStop}%
\bibitem [{\citenamefont {Suris}(1990)}]{suris1990hamiltonian}%
  \BibitemOpen
  \bibfield  {author} {\bibinfo {author} {\bibfnamefont {Y.~B.}\ \bibnamefont
  {Suris}},\ }\href@noop {} {\bibfield  {journal} {\bibinfo  {journal}
  {Matematicheskoe modelirovanie}\ }\textbf {\bibinfo {volume} {2}},\ \bibinfo
  {pages} {78} (\bibinfo {year} {1990})}\BibitemShut {NoStop}%
\bibitem [{\citenamefont {Boris}(1970)}]{boris70}%
  \BibitemOpen
  \bibfield  {author} {\bibinfo {author} {\bibfnamefont {J.}~\bibnamefont
  {Boris}},\ }in\ \href@noop {} {\emph {\bibinfo {booktitle} {Proceedings of
  the Fourth Conference on Numerical Simulation of Plasmas}}}\ (\bibinfo
  {publisher} {Naval Research Laboratory, Washington D. C.},\ \bibinfo {year}
  {1970})\ p.~\bibinfo {pages} {3}\BibitemShut {NoStop}%
\bibitem [{\citenamefont {Ware}(1970)}]{Ware_1970}%
  \BibitemOpen
  \bibfield  {author} {\bibinfo {author} {\bibfnamefont {A.~A.}\ \bibnamefont
  {Ware}},\ }\href {\doibase 10.1103/physrevlett.25.15} {\bibfield  {journal}
  {\bibinfo  {journal} {Physical Review Letters}\ }\textbf {\bibinfo {volume}
  {25}},\ \bibinfo {pages} {15} (\bibinfo {year} {1970})}\BibitemShut {NoStop}%
\bibitem [{\citenamefont {Hairer}(1999)}]{hairer1999backward}%
  \BibitemOpen
  \bibfield  {author} {\bibinfo {author} {\bibfnamefont {E.}~\bibnamefont
  {Hairer}},\ }\href {\doibase 10.1007/s002110050469} {\bibfield  {journal}
  {\bibinfo  {journal} {Numerische Mathematik}\ }\textbf {\bibinfo {volume}
  {84}},\ \bibinfo {pages} {199} (\bibinfo {year} {1999})}\BibitemShut
  {NoStop}%
\end{thebibliography}%

\end{document}